\documentclass[preprint2]{aastex6}

\usepackage{booktabs}
\renewcommand\plotone[1]{\includegraphics[width=\linewidth]{#1}}

\newcommand\kmsec{{\rm km~s^{-1}}}
\newcommand\vtan{v_{\rm tan}}
\newcommand\mbol{M_{\rm bol}}

\newcommand\teff{T_{\rm eff}}
\newcommand\K{\rm K}
\newcommand\logg{{\rm log}~g}
                                                 
\shorttitle{White Dwarf Luminosity Function}
\shortauthors{Munn et al.}

\begin{document}

\title{A Deep Proper Motion Catalog Within the Sloan Digital Sky Survey Footprint. II. The White Dwarf Luminosity Function}

\author{Jeffrey A. Munn and Hugh C. Harris}
\affil{US Naval Observatory, Flagstaff Station, 10391 W. Naval Observatory
Road, Flagstaff, AZ 86005-8521; jam@nofs.navy.mil}

\author{Ted von Hippel}
\affil{Center for Space and Atmospheric Research, Embry-Riddle Aeronautical University,
Daytona Beach, FL 32114-3900}

\author{Mukremin Kilic}
\affil{University of Oklahoma, Homer L. Dodge Department of Physics and
Astronomy, 440 W. Brooks Street, Norman, OK 73019}

\author{James W. Liebert}
\affil{University of Arizona, Steward Observatory, Tucson, AZ 85721}

\author{Kurtis A. Williams}
\affil{Department of Physics and Astronomy, Texas A\&M University--Commerce,
P.O. Box 3011, Commerce, TX 75429}

\author{Steven DeGennaro}
\affil{Department of Astronomy, University of Texas at Austin,
  1 University Station C1400, Austin, TX 78712-0259}

\author{Elizabeth Jeffery}
\affil{BYU Department of Physics and Astronomy, N283 ESC, Provo, UT 84602}

\author{Kyra Dame and A. Gianninas}
\affil{University of Oklahoma, Homer L. Dodge Department of Physics and
Astronomy, 440 W. Brooks Street, Norman, OK 73019}

\and

\author{Warren R. Brown}
\affil{Smithsonian Astrophysical Observatory, 60 Garden St., Cambridge, MA 02138}

\begin{abstract}
A catalog of 8472 white dwarf (WD) candidates is presented, selected using reduced proper motions from the deep proper motion catalog of \citet{munn2014}.  Candidates are selected in the magnitude range $16 < r < 21.5$ over 980 square degrees, and $16 < r < 21.3$ over an additional 1276 square degrees, within the Sloan Digital Sky Survey (SDSS) imaging footprint.  Distances, bolometric luminosities, and atmospheric \replaced{composition}{compositions} are derived by fitting SDSS $ugriz$ photometry to pure hydrogen and helium model atmospheres\added{ (assuming surface gravities $\logg = 8$)}.  The disk white dwarf luminosity function (WDLF) is constructed using a sample of 2839 stars with $5.5 < \mbol < 17$, with statistically significant numbers of stars cooler than the turnover in the luminosity function.  The WDLF for the halo is also constructed, using a sample of 135 halo WDs with $5 < \mbol < 16$.  We find space densities of disk and halo WDs in the solar neighborhood of $5.5 \pm 0.1 \times 10^{-3}\ {\rm pc}^{-3}$ and $3.5 \pm 0.7 \times 10^{-5}\ {\rm pc}^{-3}$, respectively.  We resolve the bump in the disk WDLF due to the onset of fully convective envelopes in WDs, and see indications of it in the halo WDLF as well.
\end{abstract}


\section{Introduction}

White dwarfs (WD) are the endpoint of stellar evolution for stars lighter than 8 -- 10~$M_\sun$
\citep{williams2009, garcia-berro2016}, or greater than 97\% of Galactic stars.
As the direct remnants of earlier star formation, WDs are an important tool in studying the evolution of our Galaxy.
The basic observable when studying star formation history with WDs is the distribution of WDs in luminosity, or the luminosity function (LF; see \citealt{garcia-berro2016} for a recent review of both the observational and theoretical work on the white dwarf luminosity function).
In particular, for a given stellar population, the location and shape of the peak and turnover in the LF at the faint end can be used to constrain the age of that population \citep{liebert1979, winget1987}.

The bright end ($\mbol \la 13$) of the white dwarf luminosity function (WDLF) is populated by hot WDs whose optical colors are distinct from other stellar populations, allowing clean samples of hot WDs to be found based on photometry alone.  Thus, even the earliest LFs for hot WDs contained hundreds of stars, including those produced from the Palomar-Green   \citep{green1980, fleming1986, liebert2005, bergeron2011} and Kiso \citep{ishida1982, wegner1994, limoges2010}  ultraviolet excess surveys.
LFs generated from modern large-scale spectroscopic surveys have been based on samples of thousands of hot WDs, including those from the SDSS \citep{hu2007, degenarro2008, krzesinski2009} and the Anglo-Australian 2dF QSO Redshift Survey \citep{vennes2002, vennes2005}.  Thus, the bright end of the WDLF is defined with high statistical significance.

The majority of WDs however are fainter, with colors indistinguishable from subdwarfs, making their selection using photometry alone impossible.  The first study to resolve the peak of the LF and obtain samples of WDs fainter than the turnover was \citet{liebert1988}, based on a sample of 43 cool WDs selected from the Luyten Half-Second Catalog \citep{luyten1979}  to have $M_v > 13$ (and later reanalyzed with additional spectroscopy and photometry by \citealt{leggett1998}).  Subsequent proper-motion based studies had similar sample sizes \citep{evans1992, oswalt1996, knox1999}.
The first major increase in sample size was that of \citet[][hereafter H06]{harris2006}, which used Sloan Digital Sky Survey \citep[SDSS;][]{york2000,gunn1998,gunn2006,fukugita1996} photometry and proper motions from a combined catalog \citep{munn2004, munn2008} of SDSS and USNO-B \citep{monet2003} astrometry to generate a sample of 6000 reduced-proper-motion selected WDs.  \citet[][hereafter RH11]{rowell2011} similarly used reduced proper motions to select 10,000 WDs from the SuperCOSMOS sky survey \citep{hambly2001a, hambly2001b, hambly2001c}.  Both surveys provide a Galactic disk WDLF with high statistical significance in the luminosity range $6 < \mbol < 15$, and clearly define the peak of the disk WDLF at $\mbol \sim 15$.  However, neither provides many stars fainter than the turnover, with only 4 stars  in H06 and 48 in RH11 with $\mbol > 15.5$ (in their $\vtan > 30~\kmsec$ samples).  Both surveys are dependent on the classic Schmidt telescope photographic surveys for one or both epochs, and thus are limited to the depth of the photographic surveys, $r \sim 19.5$.  A hydrogen atmosphere WD one magnitude fainter than the turnover has $M_r \sim 16$, which at a faint limit of $r \sim 19.5$ corresponds to only 50~pc.  Thus the small surveyed volume severely limits the number of WDs detectable past the turnover.  RH11 detected more such WDs as their sky coverage is nearly six times as great as that for H06 (roughly 30,000~deg$^2$ for RH11 versus 5300~deg$^2$ for H06).  Both papers also present a LF for clean samples of Galactic halo WDs, defined kinematically by requiring $\vtan > 200~\kmsec$.  H06 detect only 18 such halo WDs, while RH11 with their greater sky coverage detect 93.  The sample sizes of halo WDs are limited by the much smaller density of halo stars within the solar neighborhood.  The RH11 sample is by far the largest sample of halo WD candidates to date.  Note that, unlike many of the other samples discussed above, most of the WDs in the H06 and RH11 samples lack spectroscopic confirmation, though contamination by non-WDs is thought to be both understood and small \citep{kilic2006, kilic2010a}.

Recent work has concentrated on producing nearly complete samples of WDs in the local volume.  LFs have been produced using WDs within 20 \citep{giammichele2012}, 25 \citep{holberg2016}, and 40~pc \citep{limoges2015, torres2016} of the Sun, with sample sizes of 147, 232, and 501 WDs, respectively.  The 40~pc LF has 22 stars fainter than the turnover ($\mbol > 15.5$).  No halo WDs were found in any of the samples, with three possible candidates in the 40~pc sample.

In order to address the paucity of both disk WDs fainter than the turnover and halo WDs in our earlier work in H06, in 2009 we started a survey to re-observe parts of the SDSS imaging footprint, obtaining a second epoch which, combined with SDSS astrometry, would yield proper motions roughly 2 magnitudes fainter than that obtainable using the Schmidt surveys \citep{munn2014}.
This paper presents a catalog of WD candidates selected from that survey, and the resultant disk and halo WDLFs.  Individual objects with additional follow-up observations have been presented in previous papers \citep{kilic2010b, dame2016}.  Section~\ref{section-sample} describes the sample selection. Section~\ref{section-model-fits} describes the fitting of atmospheric models to SDSS photometry to derive distances, effective temperatures, and bolometric luminosities for the sample WDs.  Section~\ref{section-luminosity-function} presents the luminosity functions, Section~\ref{section-catalog} presents the catalog, and Section~\ref{section-summary} summarizes our results.

\section{Sample Selection}
\label{section-sample}

\subsection{Sky Coverage}

The WD sample is drawn from the deep proper motion survey of \citet[][hereafter M2014]{munn2014}.   This paper uses only the data considered ``good'' from the survey, which includes 1089 square degrees of sky observed with the 90prime prime focus wide-field imager on the Steward Observatory Bok 90 inch telescope \citep{williams2004}, and an additional 1521 square degrees of sky observed with the Array Camera on the U. S. Naval Observatory, Flagstaff Station, 1.3 meter telescope.  The Bok and 1.3m data are 90\% complete to $r = 22.3$ and $r = 21.3$, respectively.

We define the term {\it field} throughout this paper as the area of sky covered by a single CCD within a single observation in M2014.  Data quality naturally varies between individual observations, due to differences in seeing, image depth, etc.  Data quality can also vary between different fields within individual observations, primarily due to the varying PSF across the large field-of-views of the Bok and 1.3m telescopes;  collimating fast, large field-of-view telescopes is not easy, and both telescopes suffered from collimation issues at times during the survey.  We thus treat each field as a separate survey, in terms of rejecting bad data, modeling the data quality, and selecting candidate WDs.

Each field images an area of sky covered by multiple SDSS scans, which typically were taken on different nights.  Thus, the epoch difference between the Bok/1.3m and SDSS data can vary for different objects within a field.  We limit our survey to fields whose minimum epoch difference is at least 3.5 years, so as to provide well measured proper motions; this reduces the area coverage by 4.2\%.  A number of fields in both the Bok and 1.3m surveys are suspect, for a variety of reasons: they appear not to obtain the depth estimated in the catalog; the image quality is poor, primarily due to poor collimation; or they have a much larger number of candidate high proper motion candidate stars than expected, indicating problems with either the image quality or astrometric calibration.  These fields are excluded from the sample, and are listed in Table~\ref{table-excluded-fields}; this reduces the area coverage by a further 0.8\%.  Image quality in the corners of the Array Camera on the 1.3m begins to deteriorate, and we find a higher incidence of false proper motions in the catalog in the corners based on visual inspection.  Thus we include only objects detected within a 0.68 degree radius of the camera center for the 1.3m data, reducing the sky coverage of the 1.3m data by 12.0\%.  We further exclude areas of sky affected by bright stars, as specified using the bright star masks given in M2014 (which are derived from those of \citealt{blanton2005}), for an additional reduction in area coverage of 2.0\%.  The sky coverage of the final sample includes 980 square degrees from the Bok survey, and 1276 square degrees from the 1.3m survey.

\begin{deluxetable}{lll}
\tablecaption{Excluded Fields\label{table-excluded-fields}}
\tablehead{
  \colhead{night\tablenotemark{a}} & \colhead{obsID\tablenotemark{b}} &
  \colhead{ccds\tablenotemark{c}}
}
\startdata
53888 &  16 & 1,2,3,4 \\
53888 &  17 & 1,2,3,4 \\
53888 &  18 & 1,2,3,4 \\
54245 &  13 & 1,2,3,3 \\
54245 &  14 & 1,2,3,4 
\enddata
\tablecomments{Table~\ref{table-excluded-fields} is published in its entirety in machine-readable format.  A portion is shown here for guidance regarding its form and content.}
\tablenotetext{a}{MJD number of the night the observation was obtained in M2014.  Corresponds to the {\it night} column in the {\it Observation Schema} (Table~2) in M2014.}
\tablenotetext{b}{Observation number in M2014, unique within a given night.  Corresponds to the {\it obsID} column in the {\it Observation Schema} (Table~2) in M2014.}
\tablenotetext{c}{List of CCDs for this observation whose data were suspect, and thus excluded.}
\end{deluxetable}

\subsection{Clean Star Sample}

We start with a clean sample of SDSS stars in the $r$ band within our fields, by requiring: (1) that they pass the set of criteria suggested on the SDSS DR7 Web site for defining a clean sample of point sources in the $r$ band\footnote{
  \url{http://www.sdss.org/dr7/products/catalogs/flags.html}}; and (2) that they not be considered a moving object within a single SDSS observation (e.g., an asteroid), according to the criteria adopted for the SDSS Moving Object Catalog\footnote{\url{http://www.astro.washington.edu/users/ivezic/sdssmoc/sdssmoc1.html}} \citep{ivezic2002}.  The Bok sample is limited to stars with $16 < r < 21.5$, while the 1.3m sample is limited to stars with $16 < r < 21.3$.  This defines the complete stellar sample from which the WD candidates will be selected.

While the proper motion of each candidate WD will be visually verified, we wish to define a relatively clean sample of stars with reliably measured proper motions via a set of cuts which reject well-defined regions of parameter space with a high contamination rate of false high proper motion objects.  To this end, we adopt the following cuts:\begin{itemize}
\item Objects must be detected by SExtractor in the Bok/1.3m surveys, and not be flagged as truncated or having incomplete or corrupted aperture data (or, for the 1.3m, being saturated).  Many of the objects detected by DAOPHOT but not SExtractor are located in the extended PSF of nearby bright stars and have unreliable centroids.
\item Objects must have a reliable DAOPHOT PSF fit in the Bok/1.3m surveys, as indicated by the number of iterations required to converge on a solution ($0 < {\rm nIter} < 10$).  The DAOPHOT centroids are used to measure the proper motion, and thus the proper motion for objects whose PSF fits failed to converge (${\rm nIter} = 10$) are unreliable.  The first two cuts, which are a measure of the
depth of M2014, reject 3.9\% and 6.8\% of the Bok and 1.3m survey objects, respectively.
\item Objects must not have a nearby neighbor, whose overlapping PSF could adversely affect the measured centroids.  For the 1.3m, objects with a neighbor within 4 arcsecs are rejected.  For the Bok survey, objects with a neighbor within $3.0 + 0.5 (23 - r_n)$ arcsecs are rejected, where $r_n$ is the $r$ magnitude of the nearby neighbor.  This cut rejects a further 3.2\% and 2.2\% of the Bok and 1.3m survey objects, respectively.
\item Objects which are not a 1-to-1 match between SDSS and the Bok/1.3m surveys, or whose difference in SDSS and Bok/1.3m $r$ magnitude exceeds 0.5 magnitudes, are rejected.  The majority of these are blends and mismatches.  This rejects 0.3\% of the remaining objects.
\end{itemize}

\begin{figure}
\plotone{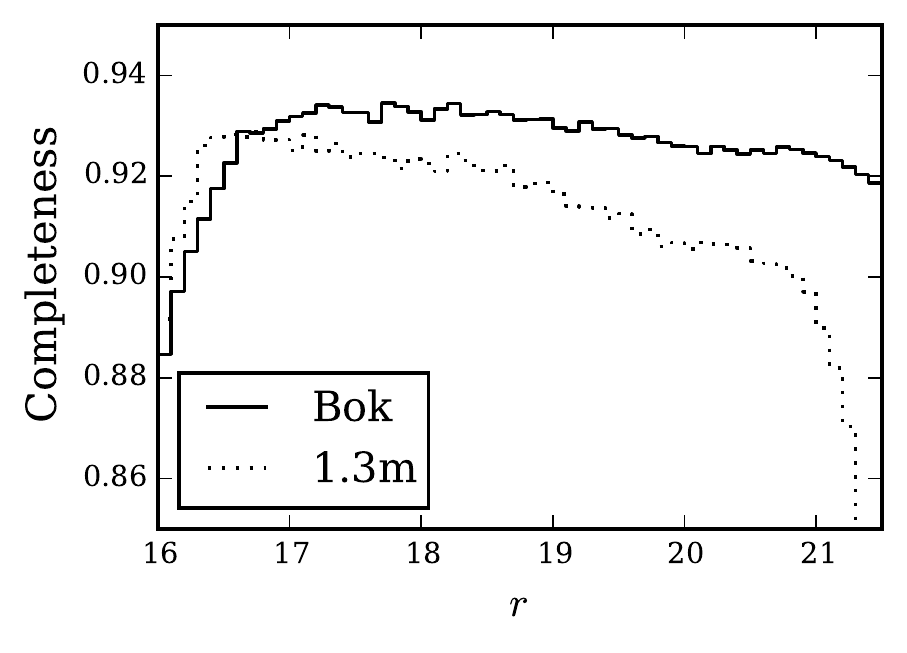}
\caption{Survey completeness versus $r$ magnitude, after applying the cuts discussed in the text.  The solid line is for the Bok survey, and the dotted line is for the 1.3m survey.}
\label{fig-completeness}
\end{figure}

These cuts reject the bulk of objects with unreliable proper motion measurements, while allowing us to define the effect on our sample completeness to allow later correction.  The survey completeness after application of these cuts is shown in Figure~\ref{fig-completeness}.

\subsection{Reduced Proper Motion Selection}

\begin{figure}
\plotone{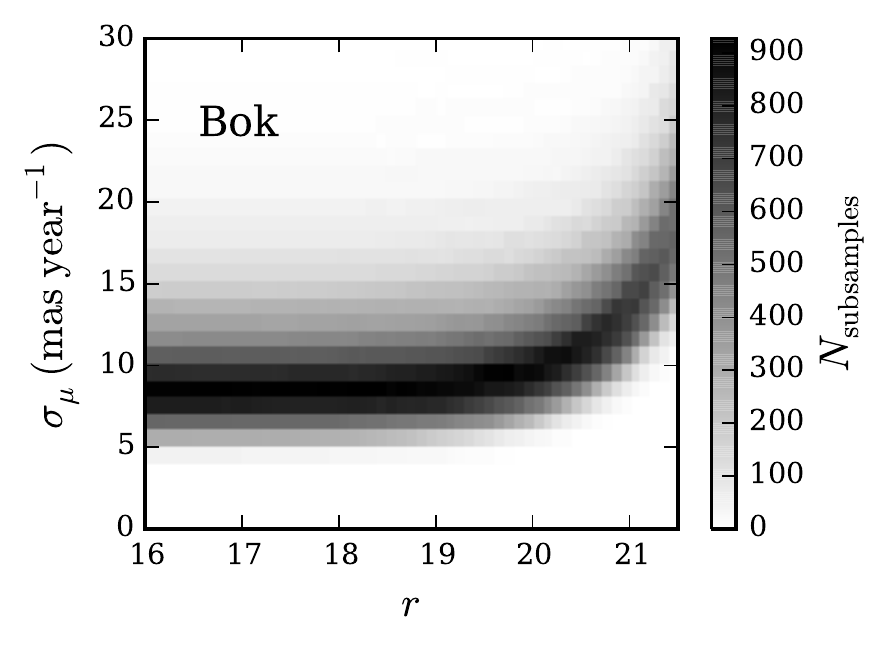}
\caption{Distribution of subsamples in proper motion error, scaled to the minimum epoch difference within their fields, versus $r$ magnitude, for the Bok survey.
  }
\label{fig-pm-errors-bok}
\end{figure}

\begin{figure}
\plotone{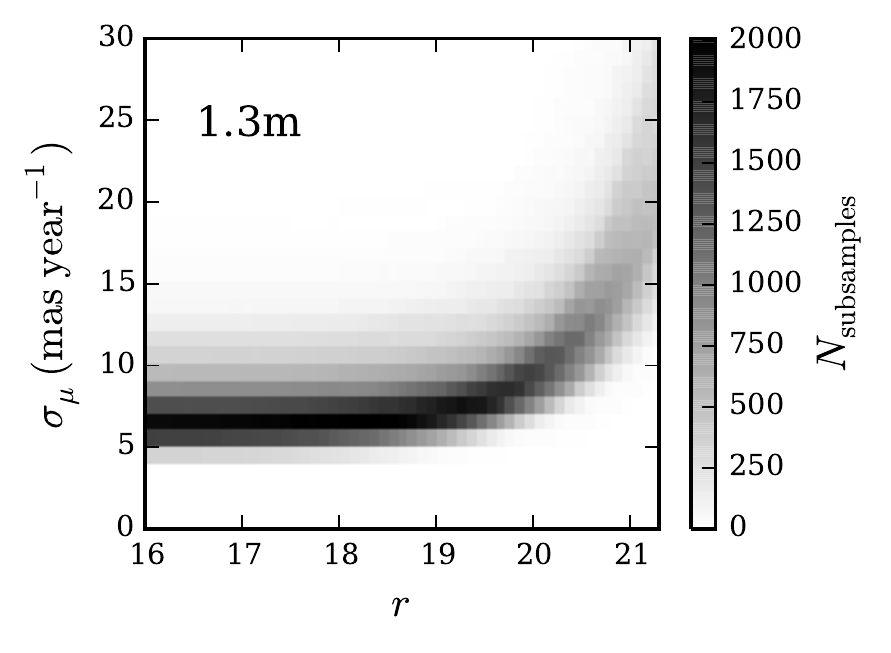}
\caption{Same as Figure~\ref{fig-pm-errors-bok}, except for the 1.3m survey.}
\label{fig-pm-errors-nofs}
\end{figure}

We select our WD candidates from stars with at least 3.5 $\sigma$ proper motions.  The proper motion errors are magnitude dependent, as well as varying between fields.  Thus, stars in each field, which we are already treating as separate samples, are further divided into magnitude bins 0.1 mag wide in $r$ (hereafter referred to as subsamples).  We calculate the mean proper motion error in each magnitude bin, scaled to the minimum epoch difference in its field, and then smooth these estimates by fitting, for each field separately, the scaled mean proper motion error versus $r$ magnitude to the function
\begin{equation}
\sigma_\mu = a + 10 ^ {0.6 (r - b)}
\end{equation}
where $r$ is the magnitude at the center of the magnitude bin and $\sigma_\mu$ is the scaled mean proper motion error in that bin.  Each magnitude bin is then treated as a separate subsample, where the proper motion error for that bin is conservatively adopted to be the proper motion error, scaled to the minimum epoch difference in the field, estimated from the fit for the faint $r$ limit of the bin.  The initial 3.5 $\sigma$ proper motion sample is then comprised of all stars with proper motions greater than 3.5 times the estimated scaled proper motion error in their subsamples.  The distribution of subsamples in scaled proper motion error versus $r$ magnitude is displayed in Figures~\ref{fig-pm-errors-bok} and \ref{fig-pm-errors-nofs}, which gives an indication of the dependence of scaled proper motion error on $r$ magnitude, and the variation in that dependency across different fields.

A common tool used to isolate cool WDs is reduced proper motion
\citep[RPM;][]{luyten1922a, luyten1922b}, defined (here in the SDSS $g$ band) as
\begin{equation}
H_g = g + 5\ {\rm log}\ \mu + 5 = M_g + 5\ {\rm log}\ \vtan - 3.379,
\end{equation}
where $\mu$ is the proper motion in arcsec~year$^{-1}$ and $\vtan$ is the tangential velocity in $\kmsec$.  Proper motion serves as a proxy for the unknown distance, as stars with similar kinematics will have similar proper motions at a given distance.  Since WDs are typically 5 -- 7 mag less luminous than subdwarfs of the same color, they are cleanly separated from subdwarfs in RPM  
versus color diagrams.  This technique was used by both H06 and RH11 (as well as numerous other studies cited in the introduction).  \citet{kilic2006, kilic2010a} obtained follow-up spectroscopy for candidate cool WDs selected from the $H_g$ versus $g-i$ diagram for the H06 sample, and found a clean separation between WDs and subdwarfs with a contamination rate of only a few percent.  Our sample will use the same selection technique, as well as the same photometric catalog (SDSS), thus their results are directly applicable to our work.

\begin{figure*}
\plotone{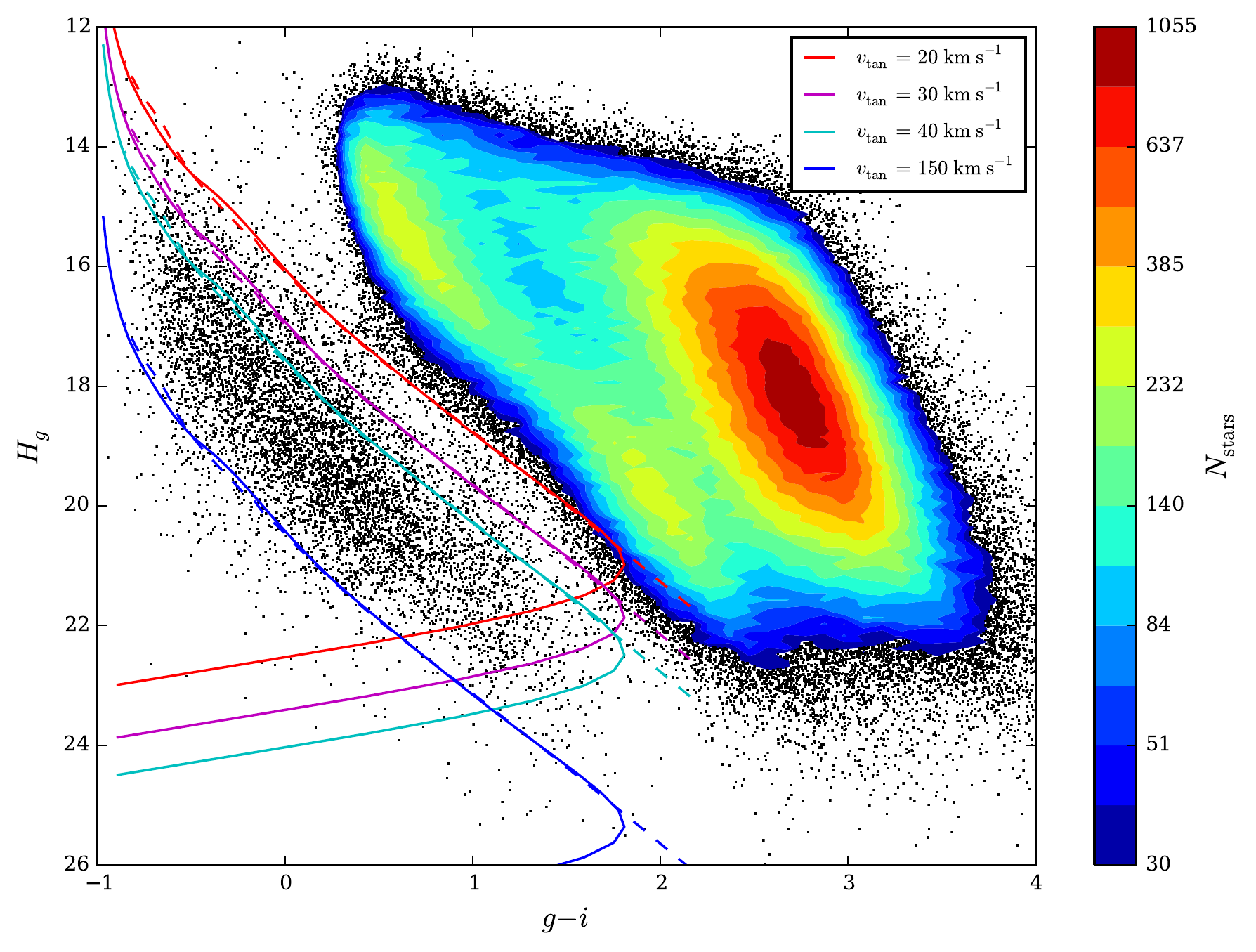}
\caption{Reduced proper motion diagram for our 3.5 $\sigma$ proper motion sample.   In the high density portion of the diagram, density contours are plotted (number of stars per bin, where each bin is 0.1 mag in $H_g$ by 0.1 mag in $g-i$).  Outside the lowest density contour, individual stars are plotted.  Evolutionary tracks for model WDs with different kinematics are plotted to indicate the expected position of WDs within the diagram.  Solid lines are pure hydrogen atmosphere WDs, while dashed lines are pure helium atmosphere WDs.  The red, magenta, cyan, and blue tracks are for WDs with tangential velocities of 20, 30, 40, and 150 $\kmsec$, respectively.  Almost all objects below the $v_{\rm tan} = 30~\kmsec$ evolutionary tracks are expected to be WDs.}
\label{fig-rpm}
\end{figure*}

The RPM diagram ($H_g$ versus $g-i$) for our 3.5 $\sigma$ proper motion sample is displayed in Figure~\ref{fig-rpm}.  In the high density portion of the diagram, density contours are plotted (number of stars per bin, where each bin is 0.1 mag in $H_g$ by 0.1 mag in $g-i$).  Outside the lowest density contour, individual stars are plotted.  The evolutionary tracks for model WDs (detailed below) of different kinematics are overlain to indicate the expected location of WDs within the RPM diagram.  Objects below the $30~\kmsec$ WD evolutionary tracks are almost exclusively expected to be WDs, with the exception of some contamination from subdwarfs at the reddest end of the evolutionary tracks for WDs with $\vtan < 40\ \kmsec$ ($g-i \sim 2$, $H_g \sim 22$). M2014 estimate a contamination rate of objects with errant proper motions of 1.5\%.  However, within the WD region of the RPM diagram, we expect a much larger contamination rate, as a 1.5\% contamination rate for the far more numerous subdwarfs can scatter a large number of subdwarfs with errant proper motions into the WD region.  Thus, all candidate WDs with $\vtan > 20\ \kmsec$ (i.e., objects below the $\vtan > 20\ \kmsec$ WD evolutionary tracks in Figure~\ref{fig-rpm}) have been examined by eye on the SDSS $r$ images, the M2014 images, and for brighter objects, on the Space Telescope Science Institute's Digitized Sky Survey scans of the photographic sky survey plates from the Palomar Oschin Schmidt and UK Schmidt Telescopes.  Of the 12,158 candidates, 3087 have errant proper motions, most due to unresolved blends with neighboring stars or image defects.  While this is a 25\% contamination rate among the WD candidates, it is only 2.1\% of the 3.5 $\sigma$ proper motion sample in the same color range, and is thus consistent with the estimate of the contamination rate in M2014.  The WD candidates with visually-determined errant proper motions are not plotted in Figure~\ref{fig-rpm}.  The remaining 9071 candidate WD candidates with $v_{\rm tan} > 20\ \kmsec$ and visually confirmed proper motions are plotted in Figure~\ref{fig-rpm}, and comprise our RPM selected sample used throughout the rest of the paper.

\section{Fits to Model White Dwarfs}
\label{section-model-fits}

We derive estimates for the distances, bolometric luminosities, effective temperatures, and atmospheric compositions for our WD candidates by fitting the SDSS photometry to the latest online set of synthetic absolute magnitudes for model WDs from P. Bergeron, G. Fontaine, P. Tremblay, and P. M. Kowalski\footnote{\url{http://www.astro.umontreal.ca/~bergeron/CoolingModels}, with models last updated 17 Oct 2011.} \citep[BFTK;][]{holberg2006, kowalski2006, tremblay2011, bergeron2011}.  It is not possible to distinguish between models with different surface gravities based on broad band photometry alone \citep{bergeron1997}.    Spectroscopic studies of WDs show a strong peak in the distribution of mass at around 0.65 -- 0.70~$M_\sun$ (depending on the details of the samples under study), corresponding to $\logg \sim 8$, with one~$\sigma$ dispersions in mass of about 0.16 -- 0.20~$M_\sun$ \citep{bergeron2001, giammichele2012, limoges2015}.  Thus, lacking surface gravity discriminators, we assume $\logg = 8$ in our fits. Each candidate is fit to both the pure hydrogen and pure helium atmosphere $\logg = 8$ model grids using variance-weighted least-squares.  Individual magnitudes which do not meet the criteria on the SDSS DR7 website for clean point sources in that filter (as used above in defining our $r$-limited stellar sample) are not used in the fit.   We also do not use individual \added{SDSS} asinh\added{\footnote{http://www.sdss.org/dr12/algorithms/magnitudes/}} magnitudes fainter than 24.02, 24.50, 24.19, 23.74, and 22.20 in $u$, $g$, $r$, $i$, and $z$, respectively, which corresponds to requiring roughly a 2~$\sigma$ detection in each filter.  The SDSS photometry is corrected to the Hubble Space Telescope flux scale, which the BFTK models use, by adding zero point offsets of -0.0424, 0.0023, 0.0032, 0.0160, and 0.0276 to the $u$, $g$, $r$, $i$, and $z$ magnitudes, respectively \citep{holberg2006}.  Errors  are estimated using the  $\chi_{\rm min}^2$ + 1 confidence boundaries in the least-squares fits.  To these are added in quadrature an estimate of the errors due to an assumed 0.3 dex scatter in $\logg$ (corresponding to a scatter in mass of $\sim 0.18~M_\sun$), derived by refitting using $\logg = 7.7$ and $\logg = 8.3$ models.  The uncertainty in $\logg$ is the dominant source of error for the bolometric luminosities and distances, leading to typical errors in the bolometric luminosities of 0.4 -- 0.5 mag.

We correct for interstellar extinction using the three dimensional reddening maps of \citet{green2015}, which provide the cumulative reddening at equally spaced distance moduli.  The reddening for each star is obtained by linearly interpolating the median reddening profile along the star's sightline.  Over half of our stars lie at distances less than the minimum distances along their sightlines considered to be reliable.  In these cases, we linearly interpolate between an assumed zero reddening at zero distance and the first reliable reddening measurement along the sightline.  Reddening is converted to extinction values in each filter using the $A_b/E(B-V)$ values from Table~6 of \citet{schlafly2011}, assuming an $R_V = 3.1$ reddening law \citep{fitzpatrick1999,schlafly2011}. Since SDSS targeted areas of low Galactic extinction, the extinction corrections are small and have little effect on the derived distances. 
The median and 90th percentile extinction in $r$ for our $\vtan > 40\ \kmsec$ sample are 0.02 and 0.06, respectively.

Each model fit is inspected by eye.  Poor fits with obvious excess flux in the $i$ and $z$ passbands are refit without the $i$ and $z$ photometry, under the assumption that the candidate WD has an unresolved M dwarf companion \citep{raymond2003, kleinman2004, smolcic2004}.  If this yields an acceptable fit, the new fit is used.  1.3\% of the sample was fit in this way.

\begin{figure}
\plotone{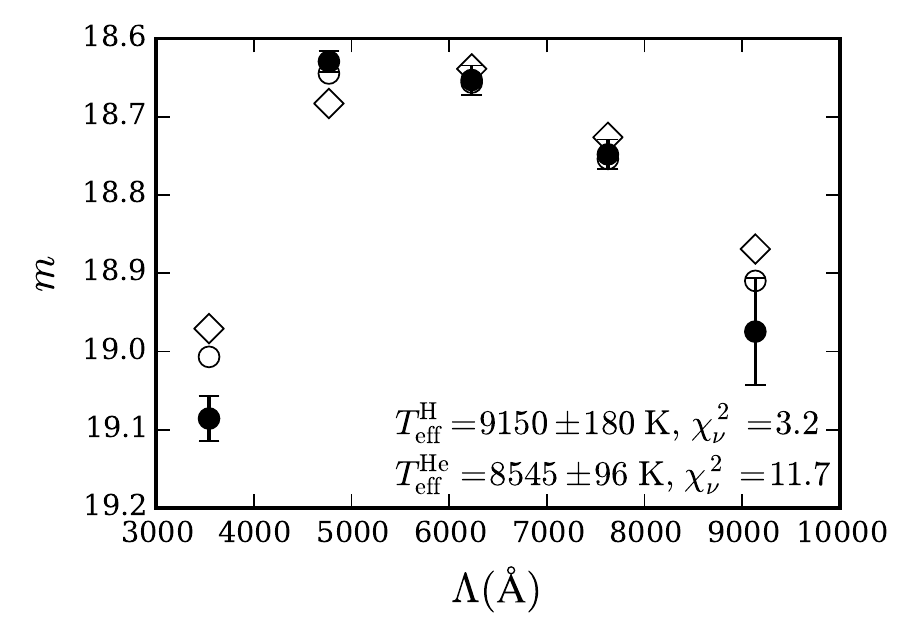}
\caption{\added{Sample white dwarf atmosphere model fit.  The filled circles with error bars display the dereddened SDSS photometry (order $ugriz$),
while the open circles and diamonds display the best fitting pure hydrogen and pure helium atmosphere model fits, respectively.  For this star, the hydrogen atmosphere model fit is considered acceptable while the helium atmosphere model fit is considered unacceptable.}}
\label{fig-sample-fit}
\end{figure}

A fit is considered acceptable if there is at least a 1\% chance of obtaining its $\chi^2_\nu$ value, and at least three SDSS magnitudes were used in the fit.  \added{A sample fit is shown in Figure~\ref{fig-sample-fit}, for which the hydrogen atmosphere model fit is considered acceptable while the helium atmosphere model fit is considered unacceptable. }599 candidates, or 7\% of the RPM-selected sample, did not have an acceptable fit for either the hydrogen or helium atmosphere model fits.  The final
WD candidate sample consists of the 8472 RPM-selected candidates with at least one acceptable fit.  For a given star, if only the hydrogen or helium fit is acceptable, then that atmospheric model is considered preferred and is used in subsequent analyses.  For stars for which both the hydrogen and helium atmosphere models yield acceptable fits, both fits are used, weighted by the expected probability of each star having a hydrogen or helium atmosphere, described below.

\begin{figure}
\plotone{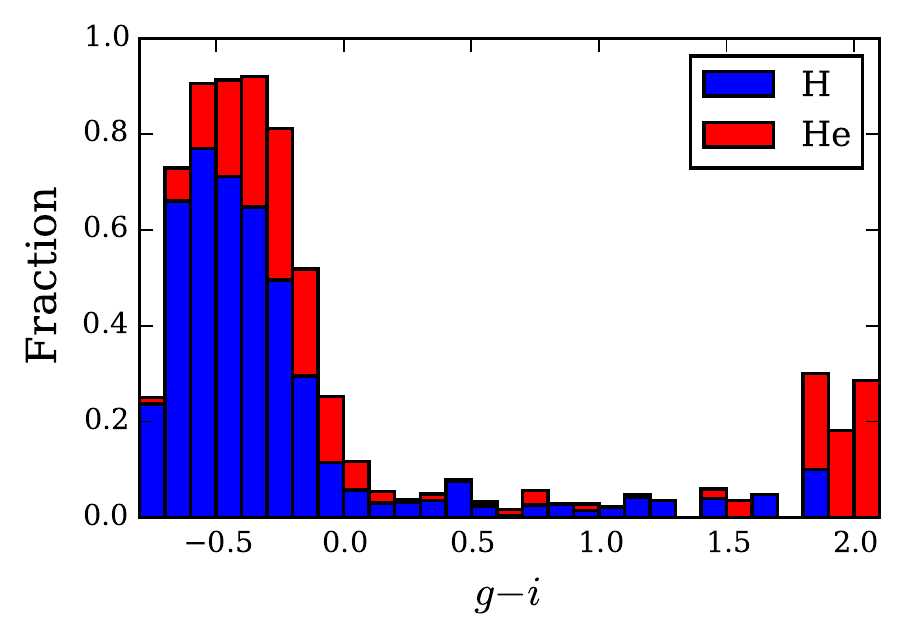}
\caption{Fraction of stars (with $\vtan > 40\ \kmsec$) for which either the pure hydrogen atmosphere model (blue histogram) or pure helium atmosphere model (red histogram) fits are considered preferred.}
\label{fig-model-fits}
\end{figure}

\begin{figure}
\plotone{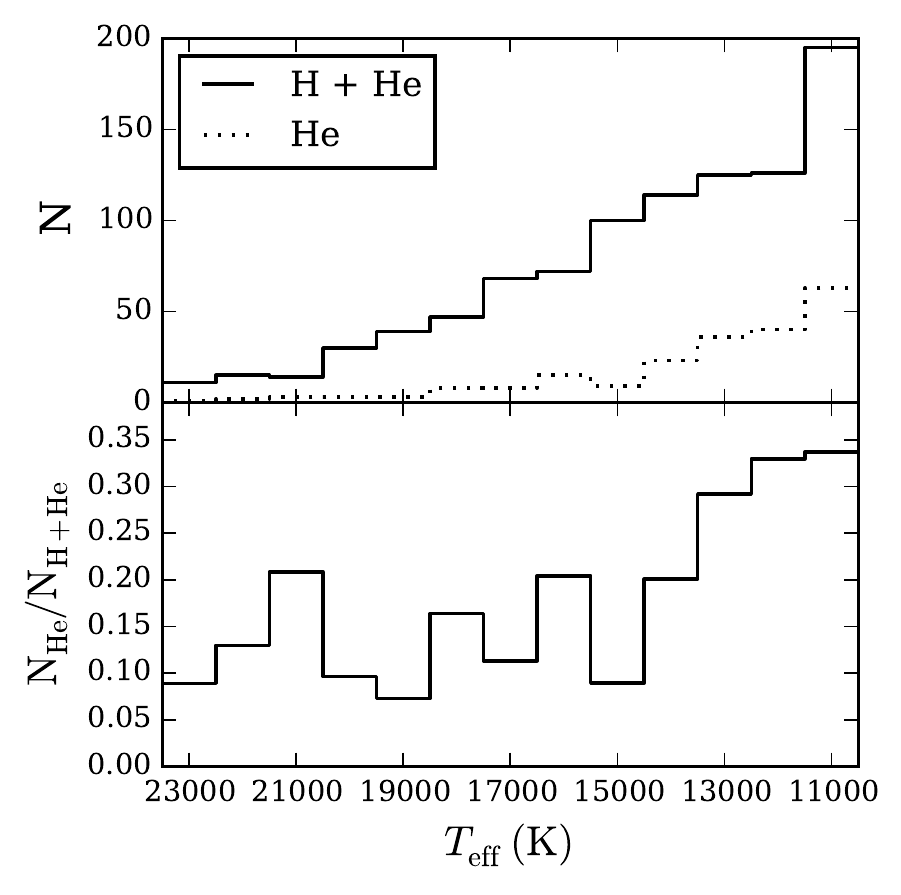}
\caption{{\it Upper panel:} Counts of stars with preferred atmospheric models versus $\teff$.  The solid line is for all stars, while the dotted line is for stars with a preferred helium atmosphere model.  {\it Lower panel:}  Fraction of stars with a preferred helium atmosphere model versus $\teff$.}
\label{fig-hot}
\end{figure}

Figure~\ref{fig-model-fits} displays the fraction of stars (with $\vtan > 40\ \kmsec$ to avoid subdwarf contamination) for which the hydrogen or helium model fits are preferred as a function of $g-i$.  82\% of stars with $g - i < -0.2$, corresponding to roughly $\teff > 10,500~\K$, have a preferred model.  For the remaining stars without a preferred model in this color range, we weight the hydrogen and helium atmosphere model fits assuming the same ratio of helium to hydrogen atmosphere WDs in a given color bin as derived from stars in that bin with preferred models.  Since the derived bolometric luminosities of WDs differ depending on whether the hydrogen or helium atmosphere model fits are used, the observed ratio of hydrogen to helium model atmospheres WDs must be adjusted to account for the difference in the maximum survey volume over which each star can be observed for the different model atmospheres.  Calculation of the maximum survey volume is described below.  Figure~\ref{fig-hot} displays the resultant helium fraction against $\teff$ for stars with  $\teff > 10,500~\K$ (based on the same sample of stars, kinematic cuts, and Galactic model used to derive our preferred disk LF, described in detail below).  The decrease in helium fraction from $\teff \sim 11,000~\K$ to $\teff \sim 15,0000~\K$ is consistent with similar trends seen in the 20 pc local volume sample of \citet[][GBD12]{giammichele2012} and 40 pc local volume sample of \citet{limoges2015}, though our results are more consistent with the overall higher helium fraction of the 20 pc sample.  For stars hotter than $\teff = 15,000~\K$, we find a helium fraction of 15\%, consistent with the estimate of $\sim$ 9\% by \citet{bergeron2011} using the Palomar-Green survey \citep{green1986}.

\begin{figure}
\plotone{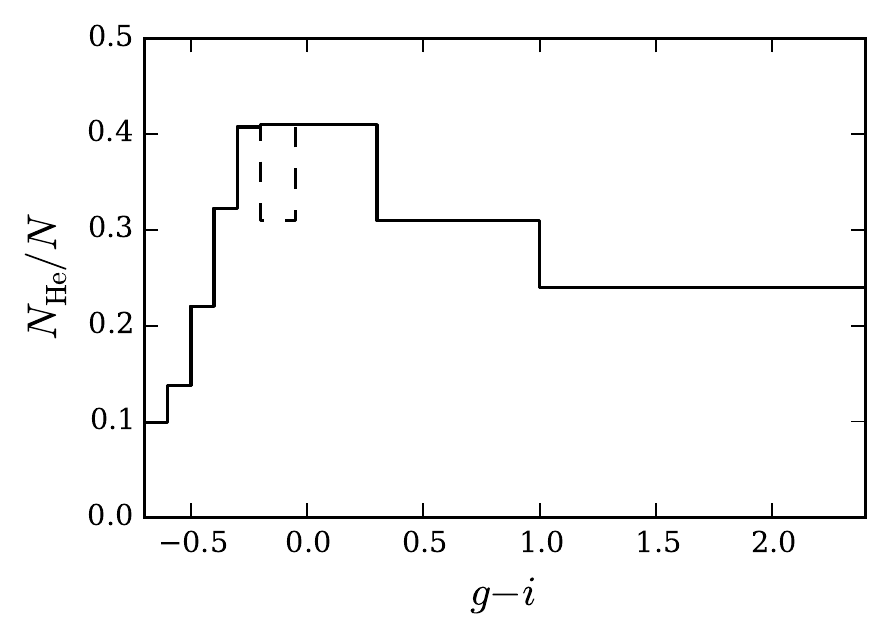}
\caption{Adopted model of helium fraction versus $g - i$ (solid line) used to weight stars lacking a preferred atmospheric model.  The model for $g - i < -0.2$ is based on results in this paper, while those for $g - i > -0.2$ are based on the results of GBD12.  The adopted fraction in the color bin $-0.2 < g - i < 0.0$ has been inflated above the actual results of GBD12 (dashed line), so as to force a smooth model.}
\label{fig-hf}
\end{figure}

For stars with $g - i > -0.2$, most stars lack a preferred model.  In this color range, we adopt the helium
fraction versus $\teff$ results of GBD12, based on their 20 pc local volume sample.  Figure~\ref{fig-hf} displays our adopted model for helium fraction versus $g - i$, used to weight the hydrogen and helium atmosphere model fits for stars which lack a preferred atmospheric model.  Our results and those of GBD12 do not smoothly meet at the border between the two ($g - i = -0.2$).  The dashed line in the $-0.2 < g - i < 0.0$ bin indicates the actual GBD12 results in that color range.  We have chosen to inflate the helium fraction in that bin to provide a smooth match between the two data sets\deleted{, so as not to introduce an artificial feature into the luminosity function}.  The difference is certainly within the error in the estimate of the true helium fraction in that color range\added{, and has negligible impact on the resultant LFs}.

\begin{figure}
\plotone{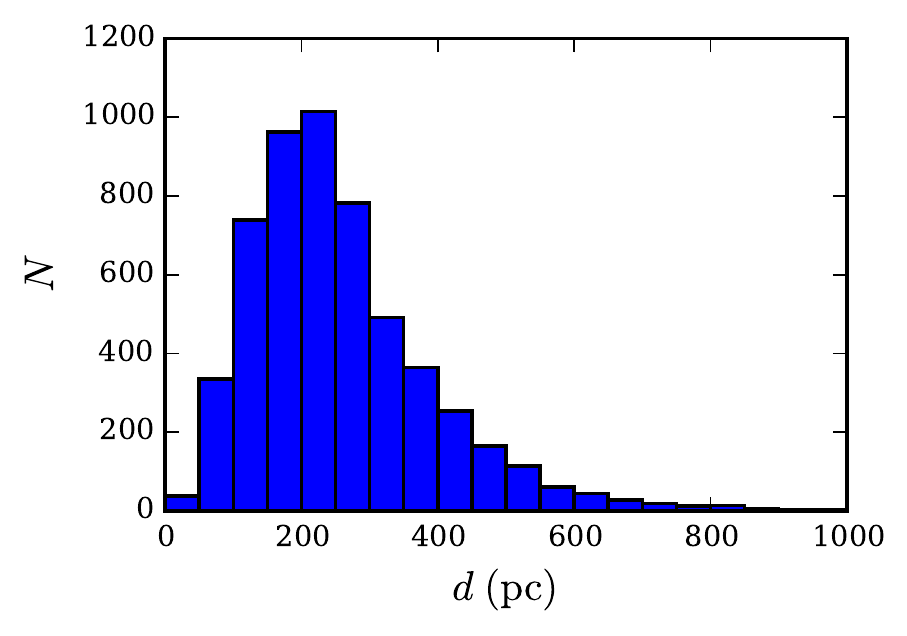}
\caption{Distribution of distances, weighted by atmospheric model fits, for the $\vtan > 40\ \kmsec$ sample (an additional 9.6 weighted stars have distances $1000 < d < 1380\ {\rm pc}$).}.
\label{fig-distances}
\end{figure}

Figure~\ref{fig-distances} shows the distribution of distances, weighted by atmospheric model fits, for the $\vtan > 40\ \kmsec$ sample (there are an additional 9.6 weighted stars with distances $1000 < d < 1380\ {\rm pc}$).  The median distance is 220~pc, while  95\% of the stars have $d < 500~{\rm pc}$.

\section{The White Dwarf Luminosity Function}
\label{section-luminosity-function}

\subsection{Method}
The WDLF is derived using a modification of the $1 / V_{max}$ method \citep{schmidt1968}, where $V_{\rm max}$ is the maximum survey volume over which each object is detectable.  Our WD sample is kinematically defined, dependent both on tangential velocity limits used to separate different Galactic populations, as well as proper motion limits which vary both between different fields and  with magnitude within each field.
We use the modified maximum survey volume of \citet[][hereafter LRH15]{lam2015} as the density estimator, which accounts for varying kinematic limits along independent lines-of-sight.
Reproducing their Equation~12, the modified maximum survey volume, calculated for each survey object independently in each survey field, is
\begin{equation} \label{eq:vmod}
V_{\rm mod} = \Omega \int_{d_{\rm min}}^{d_{\rm max}}\frac{\rho(r)}{\rho_\sun}r^2\left[\int_{a(r)}^{b(r)}P(v_{\rm tan}, r)dv_{\rm tan}\right]dr.
\end{equation}
The inner integral, referred to as the discovery fraction $\chi$, is the instantaneous fraction of objects that could be observed due to the kinematic cuts.  $P(v_{\rm tan}, r)$ is the tangential velocity ($v_{\rm tan}$) distribution, which can vary with distance, $r$, along each line-of-sight.  The limits of the integral can also vary with distance, being a combination of the tangential velocity and proper motion cuts, and are (Equations~15 and 16 from LRH15)
\begin{eqnarray}
a(r) & = & \rm{max}[v_{\rm min}, 0.00474 \mu_{\rm min}(r) r], \\
b(r) & = & \rm{min}[v_{\rm max}, 0.00474 \mu_{\rm max} r].
\end{eqnarray}
While the tangential velocity limits, $v_{\rm min}$ and $v_{\rm max}$ (in $\kmsec$), as well as the maximum proper motion cut, $\mu_{max}$ (in mas~year$^{-1}$) are constant, the minimum proper cut, $\mu_{\rm min}$, is different for each 0.1 mag wide subsample in each survey field.  All LFs presented will use a maximum proper motion cut of 1~arcsec~year$^{-1}$, for which the M2014 proper motion catalog should be complete.
Uncertainty in the discovery fraction, due to incorrect modeling of the tangential velocity distribution, represents one of the larger sources of error in luminosity functions derived from kinematically defined samples.

The outer integral in Equation~\ref{eq:vmod} represents the usual correction for varying stellar density, $\rho(r)$, along each line-of-sight, relative to the stellar density in the Galactic mid plane at the solar radius, $\rho_\sun$.  The limits of the outer integral are the distances at which each star could be observed in each survey field:
\begin{eqnarray}
d_{\rm min} & = & 10 \times 10^{(r_{\rm min} - M_r) / 5}\ {\rm pc},\\
d_{\rm max} & = & 10 \times 10^{(r_{\rm max} - M_r) / 5}\ {\rm pc},
\end{eqnarray}
where $r_{\rm min} = 16$ is the bright limit of the survey, $r_{\rm max}$ is the faint limit of the survey (21.5 for Bok fields, 21.3 for 1.3m fields), and $M_r$ is the absolute magnitude of each star, determined by the model atmosphere fits.  $\Omega$ is the solid angle subtended by each survey field.  The modified maximum survey volume for each star is then just the sum of its modified volumes in each survey field.

Stars which lack a preferred atmospheric model are included in the LF using both atmospheric model fits, weighted by the helium fraction model given in Figure~\ref{fig-hf}.  The weights are corrected for individual stars to account for the difference in maximum survey volume over which the stars could be observed using the different atmospheric model fits (H06).  Individual stars are also weighted by the probability that the star belongs to the Galactic component under study, which is a function of the adopted Galactic density and kinematic models (detailed below), and is dependent on the atmospheric model since distances differ between the hydrogen atmosphere and helium atmosphere solutions.  A correction is also applied to account for survey completeness, using a smoothed version of Figure~\ref{fig-completeness}.  The number density in a given bolometric magnitude bin is then
\begin{equation}
\Phi = \sum_{i=1}^N\frac{1}{c(r_i)} \left[ \frac{w_i^{\rm H} p_i^{\rm H}}{V_{{\rm mod},i}^{\rm H}} + \frac{w_i^{\rm He} p_i^{\rm He}}{V_{{\rm mod},i}^{\rm He}} \right],
\end{equation}
where the summation is over the stars in that bolometric luminosity bin, $c(r)$ is the survey completeness as a function of $r$ magnitude, 
$w$ are the weights assigned to the hydrogen and helium atmosphere solutions, $p$ are the probabilities the stars belong to the Galactic component under study, and $V_{\rm mod}$ are the modified maximum survey volumes.  \added{The uncertainties in the number densities are then calculated assuming Poisson statistics:
\begin{equation}
\sigma_\Phi^2 = \sum_{i=1}^N\frac{1}{c(r_i)^2} \left[ \left(\frac{w_i^{\rm H} p_i^{\rm H}}{V_{{\rm mod},i}^{\rm H}}\right)^2 + \left(\frac{w_i^{\rm He} p_i^{\rm He}}{V_{{\rm mod},i}^{\rm He}}\right)^2 \right].
\end{equation}
}

\subsection{Galactic Model}

Calculation of the modified maximum survey volume for each star requires adoption of both stellar number density and stellar kinematic models for that portion of the Galaxy covered by the survey.  We will calculate the WDLF for both the Galactic disk and halo.  No attempt will be made to calculate separate WDLFs for the thin and thick disks.  Metals sink below the photosphere in WDs due to their strong surface gravities, thus \replaced{chemical distinction between thin disk, thick disk, and halo WDs}{distinguishing between thin disk, thick disk and halo WDs based on spectroscopic chemical signatures} is not possible.  While kinematics can be used to separate disk and halo populations, the overlap in kinematics between the thin and thick disks makes separation based on kinematics difficult (though see RH11, who do present separate thin and thick disk WDLFs based on a statistical kinematic separation between thin and thick disk WDs).

We adopt the results of \citet{juric2008} to model the Galactic stellar density profile.  They applied photometric parallaxes to SDSS to model the stellar number density distribution.  They model the local ($d < 2\ {\rm kpc}$, encompassing our entire survey volume) M dwarf number distribution as the sum of two classic double exponential disks, which they interpret as the thin and thick disk.  Their preferred model gives scale heights of 300 pc and 900 pc, and scale lengths of 2600 pc and 3600 pc, for the thin and thick disks, respectively, with a local thick-to-thin disk normalization of 12\%.  We use the sum of their disk profiles as a single ``disk'' density profile.  Using stars near the main sequence turn-off, they model the Galactic halo as an oblate radial power law, with axis ratio 0.64, radial power-law index -2.77, and local halo-to-thin disk normalization of 0.51\%.

To model the kinematics of the disk, we use the results of \citet[][hereafter F09]{fuchs2009}.  They combine photometric parallaxes and proper motions to measure the first and second moments of the velocity distribution of SDSS M dwarfs in 8 slices in height above the Galactic plane, $z$, from $0 < z < 800\ {\rm pc}$ (encompassing the vast majority of our WDs).  We thus model the velocity ellipsoid as eight three-dimensional Gaussians, with the first and second moments as measured by F09, each Gaussian centered on the F09 slices ($z = 50\ {\rm pc}, 150\ {\rm pc}, ..., 750\ {\rm pc}$).  Discovery fractions are then obtained by linearly interpolating between the discovery fractions obtained from the bounding slices.
A single velocity ellipsoid is adopted for the Galactic halo, based on the results for the inner halo from \citet{carollo2010}, with dispersions ($\sigma_{v_R}, \sigma_{v_\phi}, \sigma_{v_z}$) of (150, 95, 85) $\kmsec$, and a mean rotation consistent with zero.
The velocity ellipsoids in Galactic cylindrical coordinates are projected onto the tangent plane following \cite{murray1983}.

\begin{figure}
\plotone{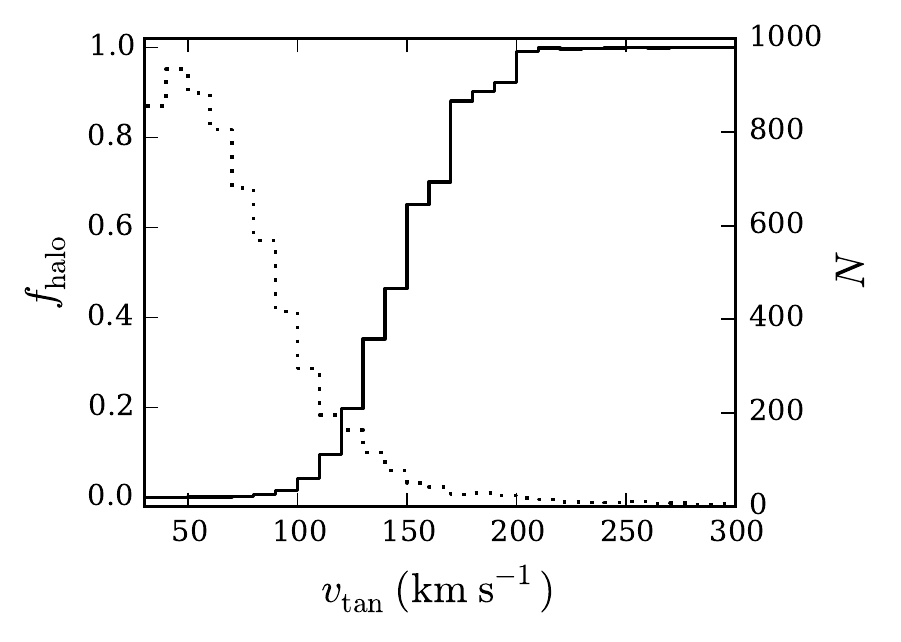}
\caption{{\it Solid line (left axis):} Expected fractional contribution of halo stars versus tangential velocity.  {\it Dotted line (right axis):} Total number of WD candidates versus tangential velocity.  An additional 47.4 (weighted) stars have $\vtan > 300\ \kmsec$.}
\label{fig-components}
\end{figure}

Figure~\ref{fig-components} plots the expected contribution of halo stars to our sample versus tangential velocity, based on the adopted Galactic density and kinematic models.  Various cuts in $\vtan$ can be used to isolate disk and halo samples.  A clean sample of disk stars can be obtained with a $\vtan < 100\ \kmsec$ cut, while a clean sample of halo stars can be obtained by a $\vtan > 200\ \kmsec$ cut.  For the disk, a minimum cut in $\vtan$ is also required to isolate WDs from subdwarfs.

\subsection{Disk Results}

\begin{figure}
\plotone{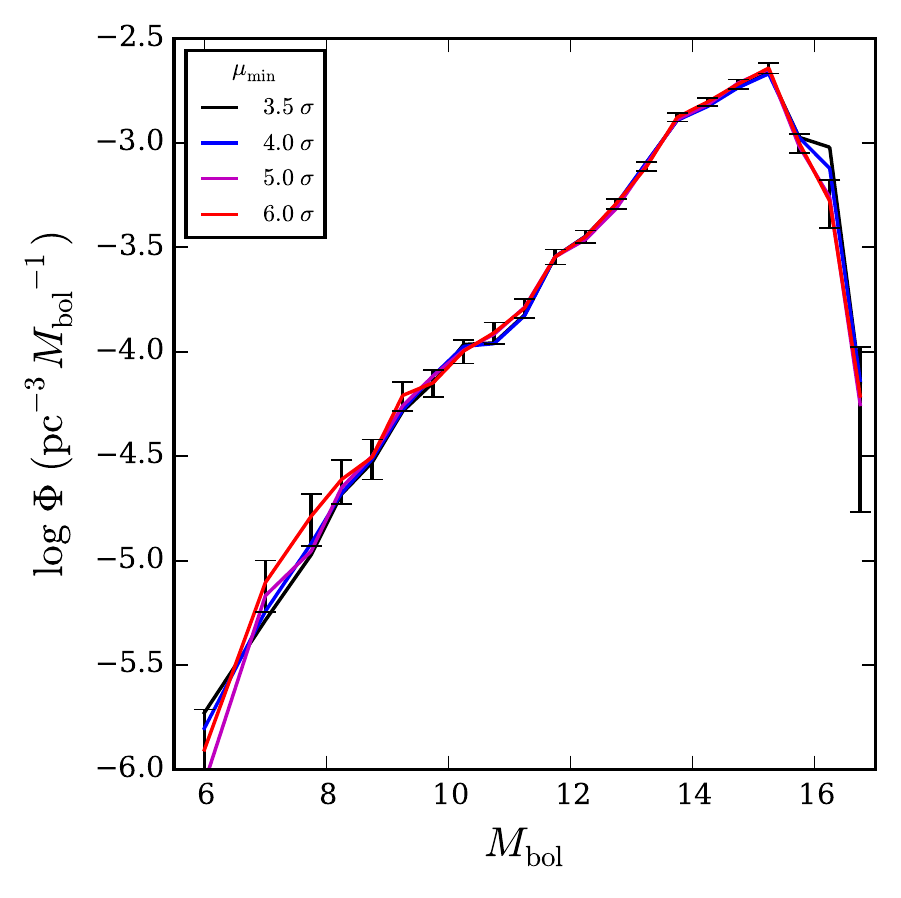}
\caption{LFs for our $40 < \vtan < 120\ \kmsec$ sample, but using different lower proper motion cuts (3.5, 4, 5, and 6~$\sigma$ for the black, blue, magenta, and red curves, respectively).  Error bars are for the $\mu_{\rm min} = 6.0~\sigma$ sample.}
\label{fig-disk-pm}
\end{figure}

The size of the sample of WDs used to derive the disk WDLF is determined by the proper motion and tangential velocity cuts employed, with the trade-off that less restrictive cuts increase sample size at the risk of introducing contamination from other stellar populations.  Figure~\ref{fig-disk-pm} plots the disk LF for  samples with $40 < \vtan < 120\ \kmsec$ (this choice for isolating our disk sample is discussed below), but different lower proper motion cuts, expressed as multiples of the proper motion error, $\sigma$.  Using 3.5, 4, 5, and 6~$\sigma$ lower proper motions cuts yields samples of 4736, 3944, 2839, and 2135 stars, respectively.  All four LFs agree within the errors, except for the region fainter than the turnover at $\mbol \sim 15$, where the WD density in the 3.5 and 4~$\sigma$ samples are elevated relative to the 5 and 6~$\sigma$ samples.  This is likely due to the scattering of subdwarfs with large proper motion errors into the WD region of the RPM diagram.  Since we are particularly interested in the faint end of the LF, we will conservatively adopt the $5~\sigma$ sample for our preferred disk sample.

\begin{figure}
\plotone{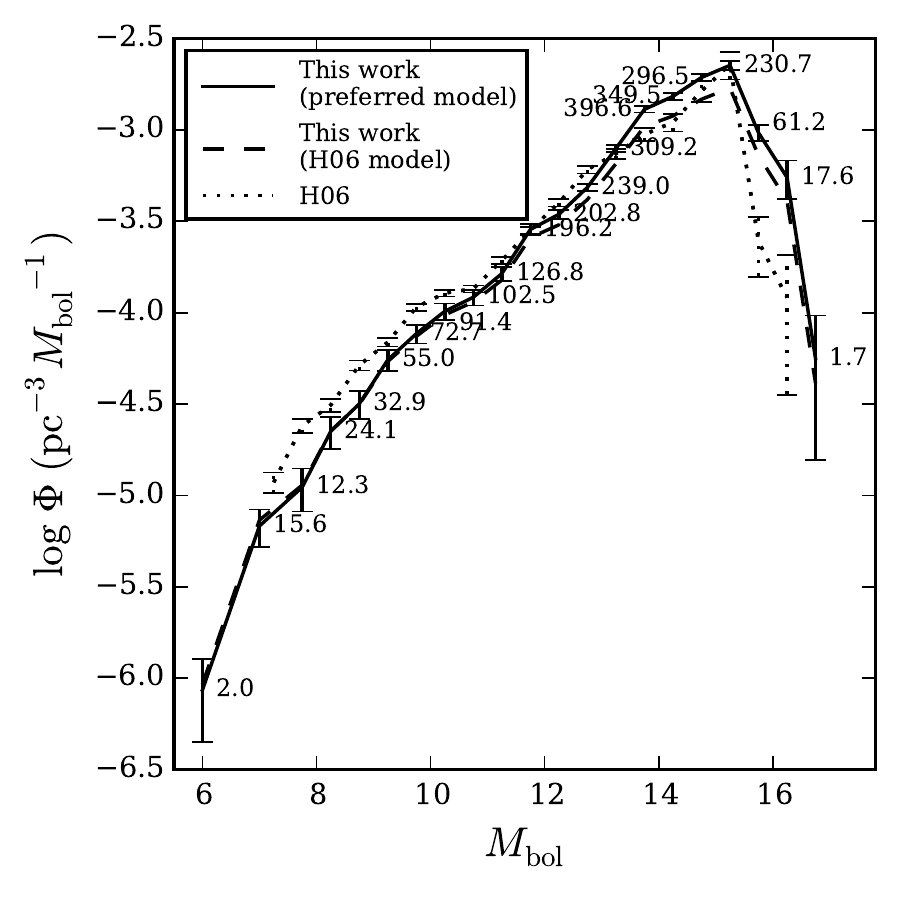}
\caption{{\it Solid line:} Preferred disk WDLF, using the $40 < \vtan < 120\ {\rm \kmsec}$ 5~$\sigma$ sample and our preferred Galactic model.  Luminosity bins are 1 mag wide from $5.5 < \mbol < 7.5$, and 0.5 mag wide from $7.5 < \mbol < 17.0$.    Numbers next to each point are the number of weighted stars in that magnitude bin.  {\it Dotted line:} Our WDLF using the H06 Galactic density and kinematic model.  {\it Dashed line:} WDLF from H06. (The LFs are available as the Data behind the Figure).}
\label{fig-disk-lf-harris}
\end{figure}

Referring to Figure~\ref{fig-rpm}, a clean separation from subdwarfs is obtained by requiring $\vtan > 40\ \kmsec$.  Our preferred disk WDLF is based on the $5~\sigma$ sample of stars with $40 < \vtan < 120\ \kmsec$, which does introduce a small expected contamination from halo stars at the highest $\vtan$, in exchange for a larger sample (see Figure~\ref{fig-components}).  The resultant disk LF is displayed in Figure~\ref{fig-disk-lf-harris}, along with the preferred LF from H06 (their Figure~4).
Luminosity bins are 1 mag wide from $5.5 < \mbol < 7.5$, and 0.5 mag wide from $7.5 < \mbol < 17.0$.
Our LF agrees reasonably well with the H06 model in the region $11 \la \mbol \la 15$.  The dip in the LF at $\mbol \sim 11$ first seen by H06 and confirmed by RH11 is less strong in our preferred LF, though is more evident in the LF from the 3.5 and 4~$\sigma$ samples (see Figure~\ref{fig-disk-pm}).
Brighter than $\mbol \sim 10$, H06 obtain densities roughly 30\% higher than our values.
The difference between our and the H06 LFs is partly due to the different Galactic models
used to correct for variations in the Galactic density profile and velocity ellipsoid.
This is indicated in the figure by plotting the LF using our preferred sample of disk stars, but calculated using the H06 Galactic density and kinematic models.  The shape of our modified LF agrees better with the H06 LF brighter than the turnover, though with an overall offset of roughly 20\%.
The primary difference between the models as it impacts the LF is the scale height of the thin disk.  H06 used a single component disk with a scale height of 250 pc, versus the \citet{juric2008} value of 300 pc which we adopted for our thin disk component (see Figure~6 in H06 for the impact of varying the disk scale height on their LF).  Note that H06 measured a scale height of $340_{-70}^{+100}\ {\rm pc}$, but adopted 250 pc for better comparison with earlier studies.  Integrating our LF yields a total WD space density in the solar neighborhood of $5.5 \pm 0.1 \times 10^{-3}\ {\rm pc}^{-3}$, versus the H06 value of $4.6 \pm 0.5 \times 10^{-3}\ {\rm pc}^{-3}$.  Our space density using the H06 Galactic model is $4.4 \pm 0.1\ \times 10^{-3}\ {\rm pc}^{-3}$, in good agreement with the H06 value.

Our WD sample and the H06 WD sample share many of the same stars, though they were selected from different proper motion catalogs, and those stars in common use the same SDSS photometry for the WD atmosphere model fits, thus they are not entirely independent.  This is particularly true at the brighter end of the LF;  67\% of our preferred disk WD sample with $\mbol < 12$ are also in the H06 sample, while only 16\% of our sample with $\mbol > 12$ are in the H06 sample.  The overlap is less for the 3.5~$\sigma$ sample, where 55\% of the stars with $\mbol < 12$ are in the H06 sample, and only 11\% of stars with $\mbol  >12$.

\begin{figure}
\plotone{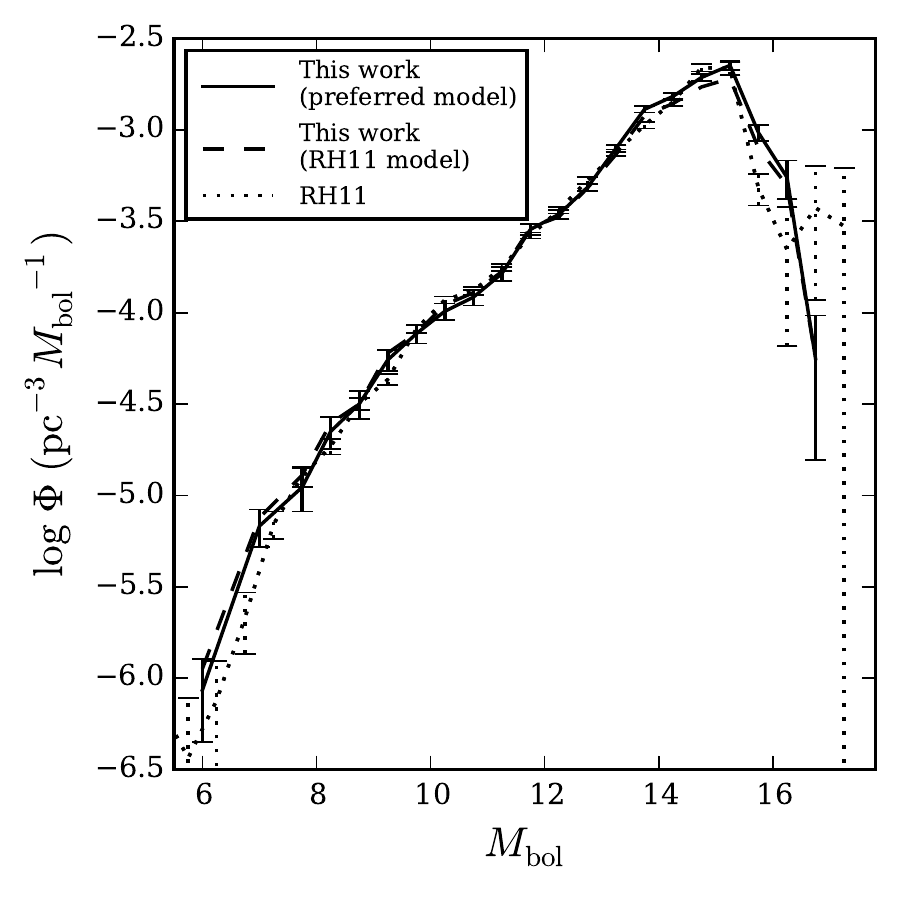}
\caption{{\it Solid line:} Preferred disk WDLF, using the $40 < \vtan < 120\ {\rm \kmsec}$ 5~$\sigma$ sample and our preferred Galactic model (same as in Figure~\ref{fig-disk-lf-harris}).  {\it Dotted line:} Our WDLF using the RH11 Galactic density and kinematic model.  {\it Dashed line:} WDLF from RH11, scaled up by a factor of 2.00. (The LFs are available as the Data behind the Figure)}
\label{fig-disk-lf-rh}
\end{figure}

Figure~\ref{fig-disk-lf-rh} again displays our preferred disk LF, but now compared to the sum of the thin and thick disk LFs from RH11 (their Figure 18).  The RH11 LF has been scaled up by a factor of 2.00 to match our LF in the region $9 < \mbol < 15.5$, consistent with their estimated incompleteness of up to 50\%.  The agreement in the shape of the LFs is better than with the H06 LF.  RH11 used a two component disk model, with thin and thick disk scale heights of 250~pc and 1500~pc, respectively, and derived fractional thin disk, thick disk, and halo contributions to the local WD density of 0.79, 0.16, and 0.05, respectively.
Using our data with the RH11 Galactic model makes for a somewhat poorer agreement with the RH11 LF.

\begin{figure}
\plotone{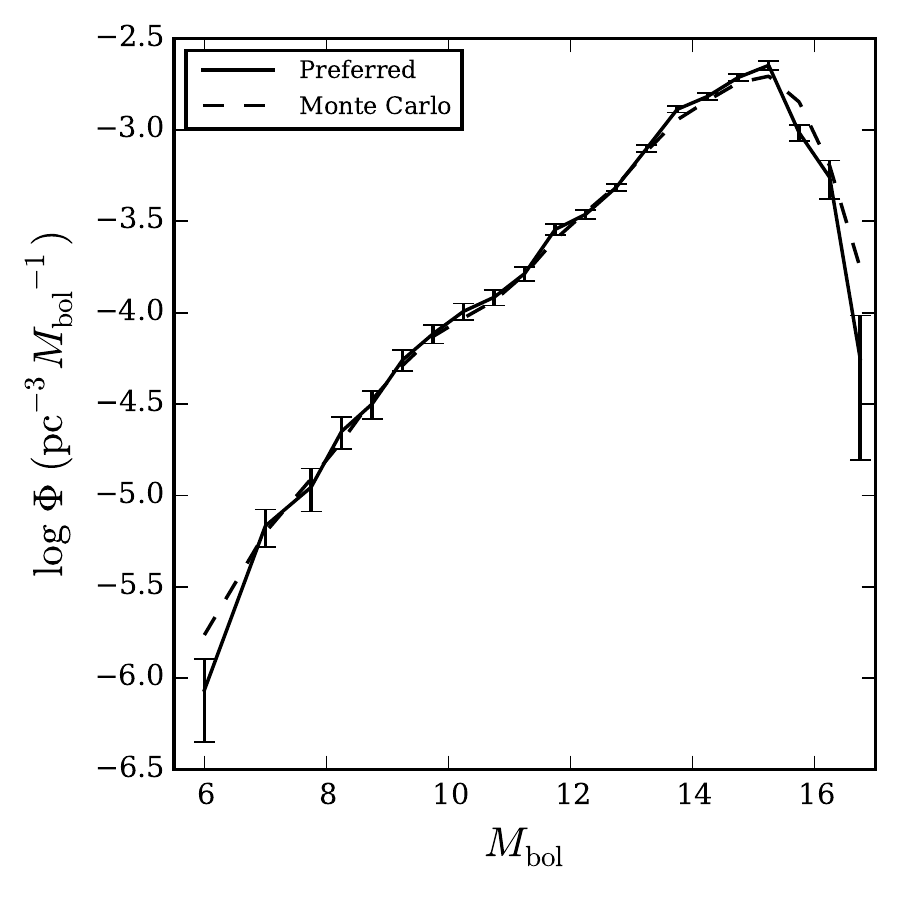}
\caption{\added{{\it Solid line:} Preferred disk WDLF (same as Figures~\ref{fig-disk-lf-harris} and \ref{fig-disk-lf-rh}).  {\it Dashed line:} Disk WDLF from a Monte Carlo simulation assuming a Gaussian dispersion in surface gravity of $\sigma_{{\rm log}~g} = 0.3$.}}
\label{fig-monte-carlo}
\end{figure}

The y-axis error bars in our LFs reflect only the Poisson errors, and do not account for other potential sources of error.  For example, at the faint end of the LF, the distribution of WD masses is poorly constrained, as the spectra are featureless and so one must rely on parallaxes, which are not available for most known WDs fainter than the turnover.  This leads to typical errors in the bolometric luminosities of around 0.5 mag, comparable to the size of the bins in our LF.
\replaced{Even if the distribution of masses is symmetrical around the peak mass, some artificial increase in the density in the luminosity bins beyond the turnover can be expected, as the sharp decline in the LF will result, for each luminosity bin, in more stars being scattered in from the adjacent brighter bin than are scattered out into the adjacent fainter bin.}
{We examine the impact on the LF of the large uncertainties in bolometric luminosities by performing a Monte Carlo simulation, in which 100 stars are generated for each star in our preferred disk LF, where $\mbol$ for each simulated star is drawn from a Gaussian distribution centered on the measured $\mbol$ with a dispersion in $\mbol$ corresponding to a dispersion in surface gravity of $\sigma_{{\rm log}~g} = 0.3$.  Figure~\ref{fig-monte-carlo} compares our preferred disk LF with the LF derived from the Monte Carlo simulation.
An increase in the density in the luminosity bins beyond the turnover can be seen, as the sharp decline in the LF results, for each luminosity bin, in more stars being scattered in from the adjacent brighter bin than are scattered out into the adjacent fainter bin.}

\begin{figure*}
\plotone{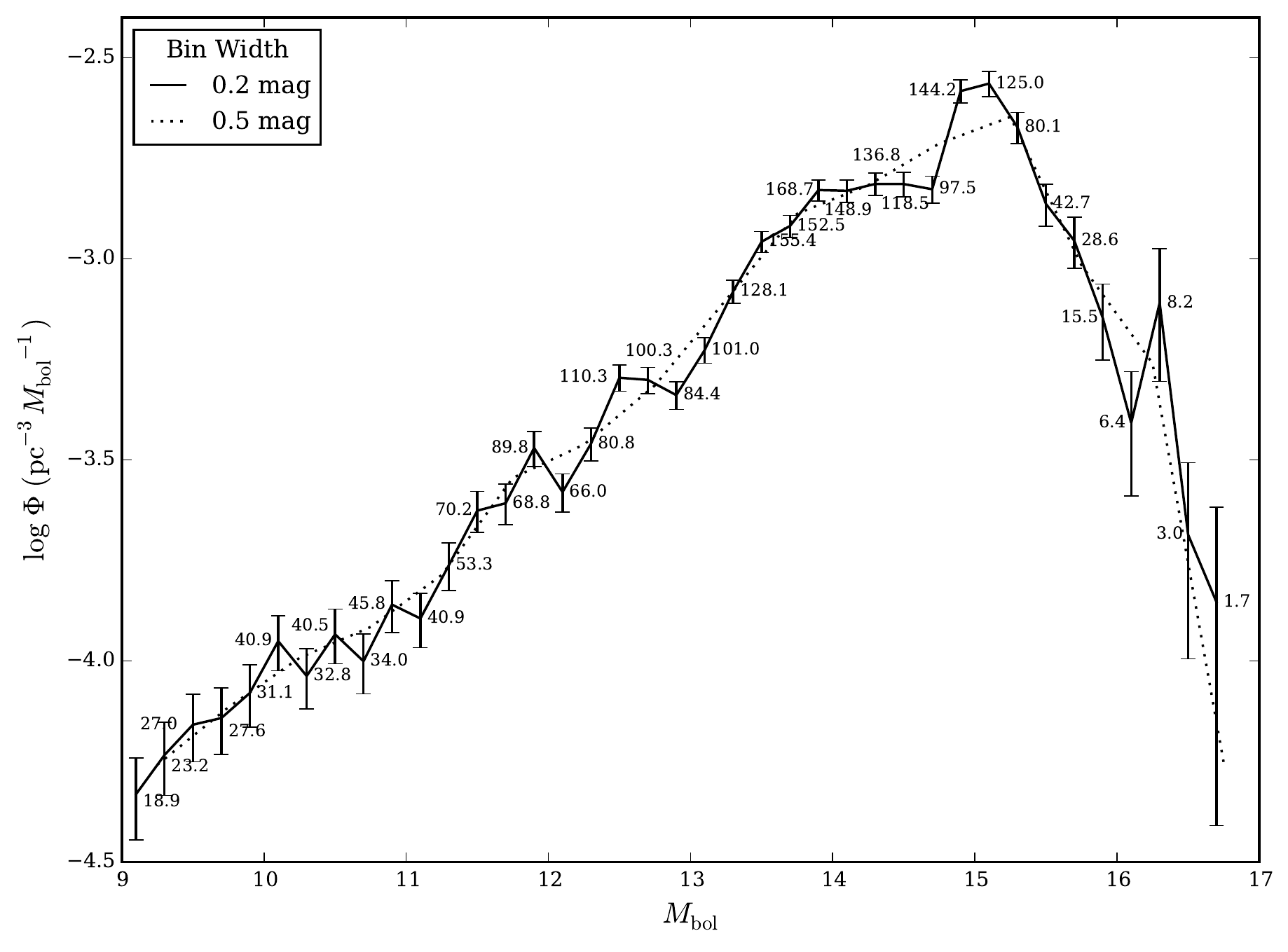}
\caption{Preferred disk LF with 0.2 mag wide bins (solid line, with numbers indicating the number of weighted stars in each bin), compared to LF with 0.5 mag wide bins (dotted line). (The LFs are available as the Data behind the Figure)}
\label{fig-disk-detail}
\end{figure*}

Figure~\ref{fig-disk-detail} displays the disk LF, but with 0.2 mag wide bins rather than 0.5 mag (limited to $\mbol > 9$, as there are too few stars brighter than that limit to support the finer binning).  We see the same sharp rise  just before the peak ($\mbol \sim 14.6$) that H06 saw (see their Figure~9), though now with considerably greater statistical significance.  The rise also occurs about 0.2 mag brighter than in H06. H06 interpreted this feature as due to the delay in cooling that occurs when the hydrogen envelope becomes fully convective, breaking into the thermal reservoir of the degenerate core and leading to the release of excess thermal energy \citep{fontaine2001}.

\begin{figure}
\plotone{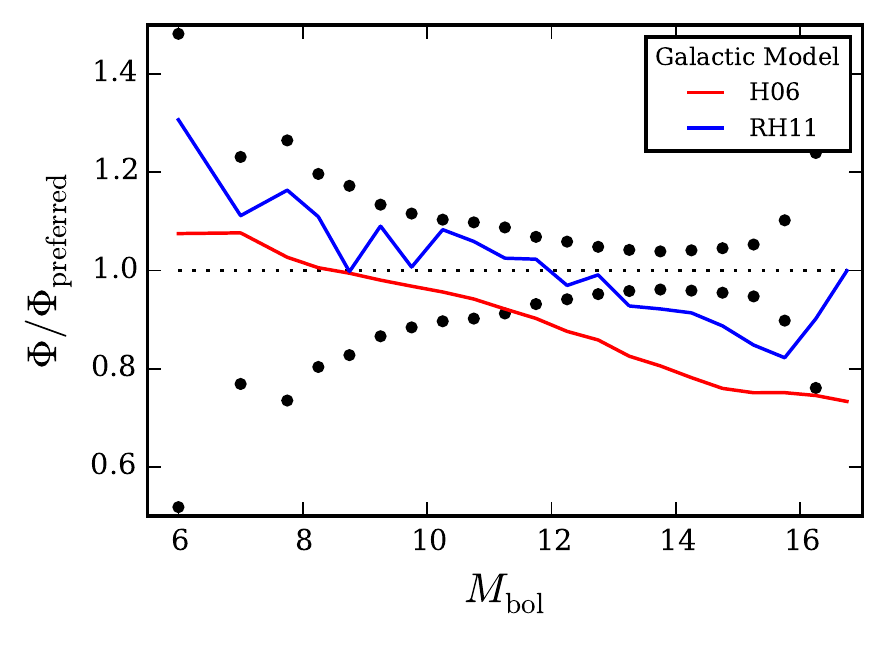}
\caption{Effect on the disk LF of different Galactic density and kinematic models.  Colored lines indicate the ratio of the LF using alternate Galactic models to our preferred LF.  The dots indicate the size of the Poisson errors in our preferred LF.}
\label{fig-disk-model}
\end{figure}

The sensitivity of the LF to the adopted Galactic model is displayed in Figure~\ref{fig-disk-model}, where the ratio of our LFs using the H06 and RH11 models to the LF using our preferred model is given.  The differences are as large as 30\%.  The dominant contributor to the differences is the different values used for the thin disk scale height.  

\begin{figure}
\plotone{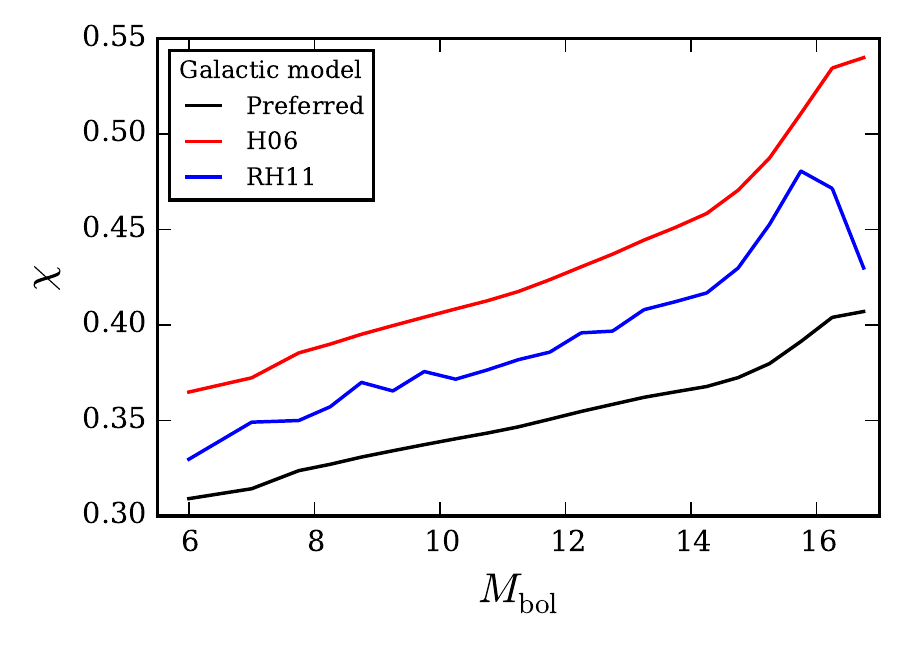}
\caption{Mean \deleted{volume-weighted }discovery fraction, $\chi$, versus $\mbol$ for our preferred disk WD sample ($40 < \vtan < 120\ \kmsec$), using different Galactic kinematic and density models.  Luminosity bins are the same as in Figure~\ref{fig-disk-lf-harris}.}
\label{fig-disk-df}
\end{figure}

Correction for the discovery fraction presents one of the larger sources of uncertainty in deriving the WDLF from kinematically defined samples.  Figure~\ref{fig-disk-df} displays the mean \deleted{volume-weighted }discovery fraction versus $\mbol$ for our preferred disk sample using three different Galactic models: our preferred model (black curve), the H06 model (red curve), and the RH11 model (blue curve).  The discovery fraction increases at fainter bolometric magnitudes because intrinsically fainter stars are on average nearer than the brighter stars, and thus have a higher expected proper motion.  The discovery fraction averaged over the entire sample for our preferred Galactic model is 0.36, thus a large correction is required.  The sensitivity of the discovery fraction to the adopted model can be seen by comparing the curves for the different models.  The H06 model yields discovery fractions typically 15\% higher than ours, though up to 30\% higher at the faint end of the LF.  H06 used a single velocity ellipsoid for their disk, whose dispersion is larger than the F09 values used in our preferred kinematic model for $|z| \la 350\ {\rm pc}$.  This leads to a higher discovery fraction, particularly at the faint end of the LF, where stars in our sample are much closer than at the bright end.  RH11 used a single velocity ellipsoid for each of their thin and thick disk components.  This yields a discovery fraction that agrees better with our preferred model, being 5\% higher at the bright end, though rising to 15\% at the faint end.  For a given sample of stars, a higher discovery fraction yields a smaller normalization correction and thereby  a lower luminosity density.

\begin{figure}
\plotone{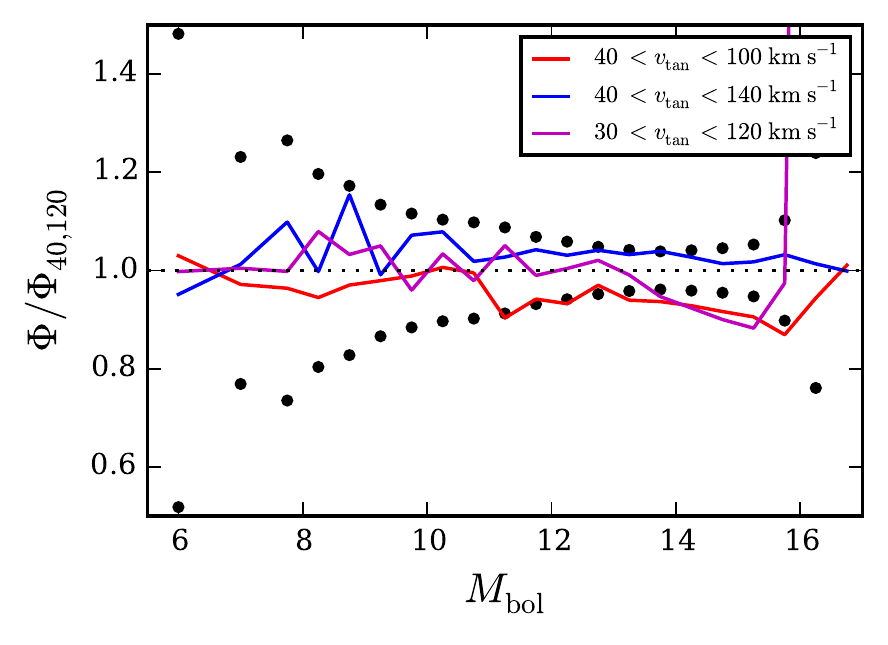}
\caption{Effect on the disk LF of different cuts in $\vtan$.  Colored lines indicate the ratio of the LF using alternate $\vtan$ cuts to our preferred LF ($40 < \vtan < 120\ \kmsec$).  The dots indicate the size of the Poisson errors in our preferred LF.}
\label{fig-disk-vtan}
\end{figure}

The accuracy of the correction for the discovery fraction can be assessed by comparing LFs using different cuts in $\vtan$.  This is done in Figure~\ref{fig-disk-vtan}, where the ratios of the LFs using $\vtan$ cuts of $40 < \vtan < 100\ \kmsec$, $40 < \vtan < 140\ \kmsec$, and $30 < \vtan < 120\ \kmsec$ to our preferred LF with $40 < \vtan < 120\ \kmsec$ are plotted.  Also plotted are the Poisson errors in our preferred LF, to allow comparison of the different sources of errors.  The large difference between the $30 < \vtan < 120\ \kmsec$ and preferred LFs for $\mbol > 15$ is due to subdwarf contamination.  Excluding that contaminated region, the models vary by typically 5 -- 10\%, comparable to the Poisson errors.

Fainter than the turnover ($\mbol > 15$) we find a higher density of stars than either H06 or RH11, though again the error bars on all three LFs are likely underestimates of the actual errors, and thus the differences are only of order 2 -- 3 $\sigma$.  In raw counts, we have 230.7 weighted stars in the peak bin of the LF, $15 < \mbol < 15.5$, versus 31 in H06 and 213 in RH11.  For $\mbol > 15.5$, we have 80.5 stars, versus 4 in H06 and 48 in RH11.  Lacking spectroscopic confirmation, some contamination from subdwarfs cannot be completely ruled out, though we expect the contamination to be small with the conservative tangential velocity and proper motion cuts we adopted.  Using follow-up spectroscopy, \citet{kilic2010a} found only one subdwarf among the 75 WD candidates with $\mbol > 14.6$ and $\vtan > 30\ \kmsec$ in the H06 sample (they find a considerably higher contamination rate for the $\vtan > 20\ \kmsec$ sample).  While we use a different proper motion catalog than H06, with only two epochs versus six in H06 and thus a higher risk of errant proper motions, we used a similar vetting procedure as H06 to confirm our proper motions, and thus we expect their results to be largely applicable to our sample.

\begin{figure}
\plotone{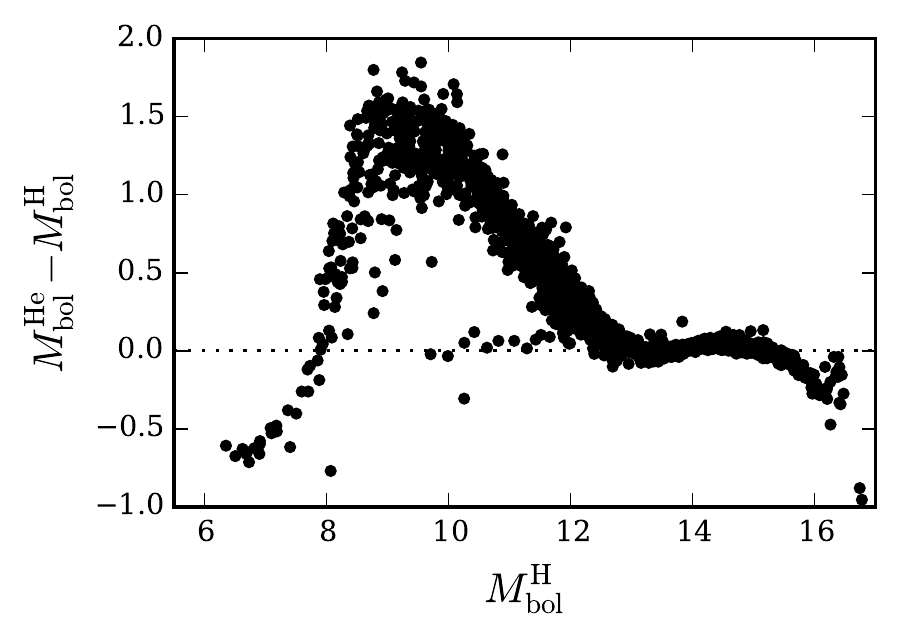}
\caption{Difference in bolometric magnitude between the hydrogen and helium model fits versus the bolometric magnitude using the hydrogen model fit for stars in our preferred disk LF.}
\label{fig-mbol-diff}
\end{figure}

\begin{figure}
\plotone{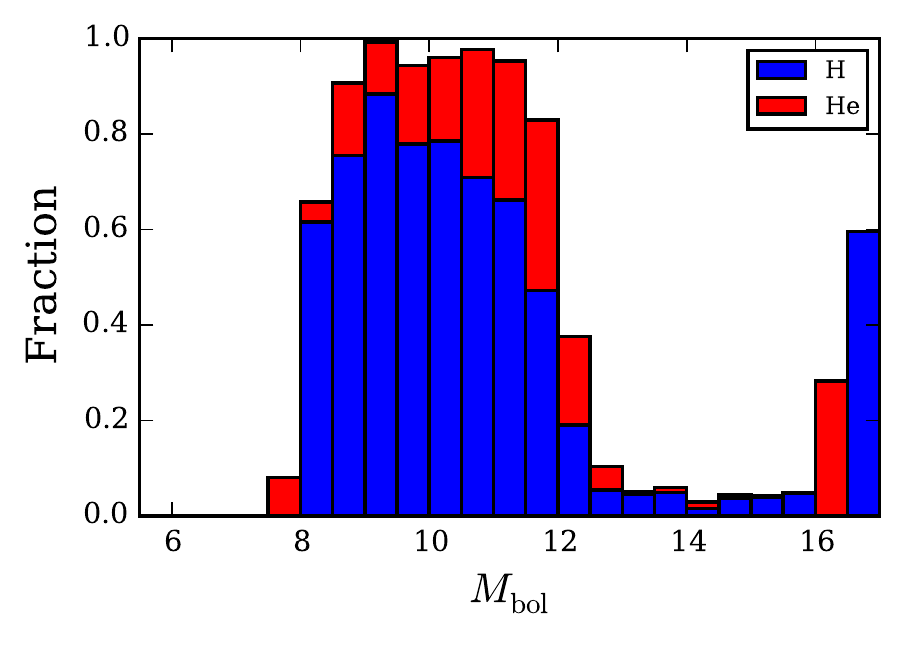}
\caption{Fraction of stars in our preferred disk sample whose fit to either the pure hydrogen atmosphere model (blue histogram) or pure helium atmosphere model (red histogram) is considered preferred as a function of bolometric luminosity.}
\label{fig-models-mbol}
\end{figure}

\begin{figure}
\plotone{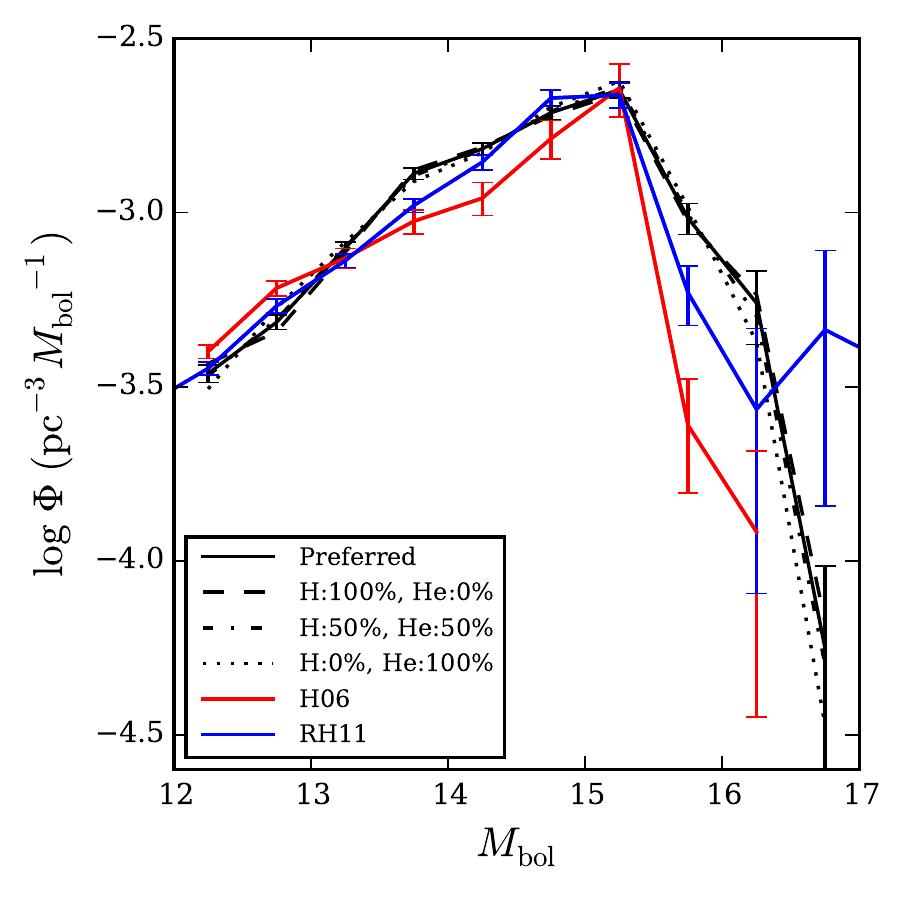}
\caption{Effect of different models for the ratio of hydrogen to helium atmosphere stars for those stars lacking a preferred atmospheric model.  {\it Solid line:} Our preferred model.  {\it Dashed line:} 100/0 H/He split.  {\it Dash-dot line:} 50/50 H/He split.  {\it Dotted line:} 0/100 H/He split. {\it Red line:} H06 LF. {\it Blue line:} RH11 LF.}
\label{fig-disk-helium}
\end{figure}

Previous studies, including both H06 and RH11, have emphasized the impact of the unknown atmospheric composition of the stars on the LF fainter than the turnover.
Figure~\ref{fig-mbol-diff} displays the difference in bolometric magnitude derived from the hydrogen and helium atmosphere model fits versus bolometric magnitude derived from the hydrogen model fits.  While the difference is large for intrinsically brighter WDs ($\mbol \la 12$, $\teff \ga 9600~{\rm K}$), optical colors allow a determination of the appropriate atmospheric composition for most of these stars due to the strong Balmer lines in DA WDs.  This is seen in Figure~\ref{fig-models-mbol}, which shows the fraction of stars in our disk sample which have a preferred atmospheric model fit, using the same binning in bolometric luminosity as used in our disk LFs.  From $12 \la \mbol \la 15.5$, just past the peak of the LF, most stars lack a preferred model, however the difference in bolometric magnitude is less than 0.1 mag, and thus has little impact on the luminosity function.  Beyond the turnover ($\mbol > 15.5$), the magnitude differences exceed 0.1 mag and increase as the intrinsic luminosity decreases.  Most stars in this luminosity range lack a preferred atmospheric model (the high fraction of stars with a preferred model in the faintest bin in Figure~\ref{fig-models-mbol} has little statistical significance, as there are only 1.7 weighted stars in this bin), and thus the uncertainty in the atmospheric composition significantly impacts the region of the LF cooler than the turnover.  This impact is indicated in Figure~\ref{fig-disk-helium}, where we plot the fainter end of the LF with different assumed helium fractions for stars which lack a preferred atmospheric model.  Our preferred model uses the helium fraction model of GDB12 in this luminosity range, which has a helium fraction of 24\% for stars fainter than the turnover.  Also plotted are the LFs of H06 and RH11, which both assumed helium fractions of 50\%.  Regardless of what helium fraction we adopt, we still find a higher density of stars beyond the turnover than either H06 or RH11.

\begin{figure}
\plotone{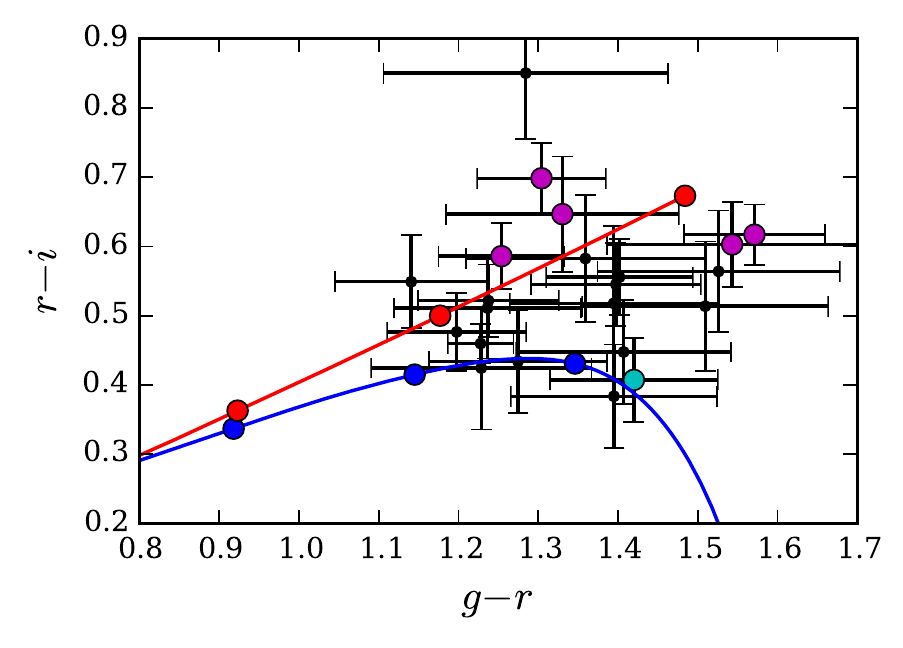}
\caption{Optical color-color plot for our preferred disk sample with $\mbol > 16$.  Stars plotted in cyan and magenta have preferred hydrogen and helium atmospheres, respectively.  The blue and red lines are the WD cooling tracks for pure hydrogen and helium atmosphere WDs, with points corresponding to $\teff$ of 4500, 4000, and 3500~K indicated (temperature decreases with increasing $g - r$ for both models).}
\label{fig-gri}
\end{figure}

\begin{figure}
\plotone{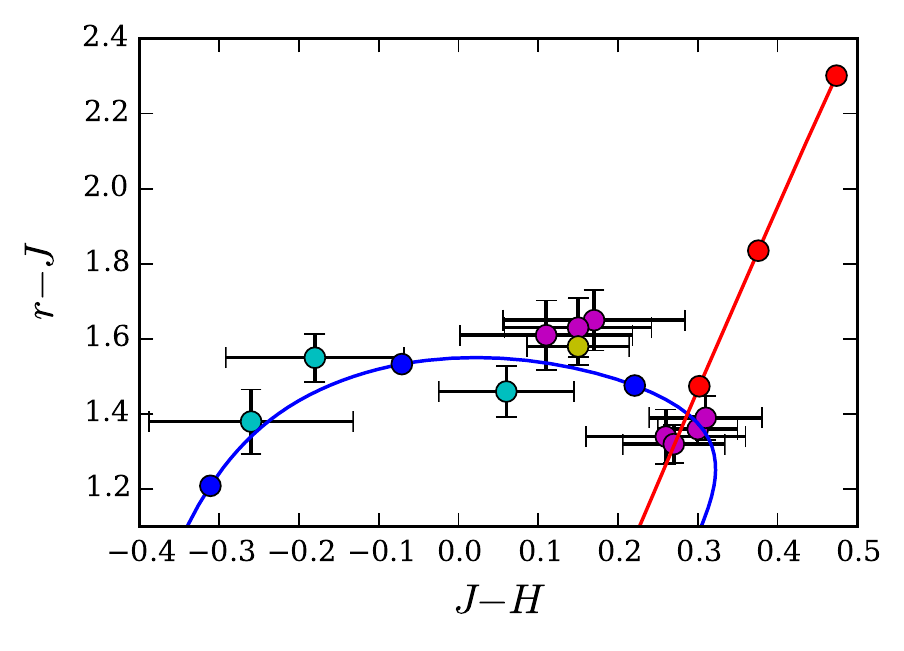}
\caption{Optical-infrared color-color plot of our preferred disk sample with $\mbol > 15.5$, for those stars with infrared photometry from D16.  Stars plotted in cyan, magenta, and yellow were classified as hydrogen, helium, and mixed atmosphere WDs by D16, respectively, however those classified as helium atmosphere WDs should be considered to not have a preferred fit (see text).  The blue and red lines are the WD cooling tracks for pure hydrogen and helium atmosphere WDs, respectively, with points corresponding to $\teff$ of 4500, 4000, and 3500~K indicated (temperature decreases with increasing $J-H$ for the helium model, but decreases with decreasing $J-H$ for the hydrogen model).}
\label{fig-rjh}
\end{figure}

Additional data can help distinguish between atmospheric models for cooler WDs.  The onset of collisionally induced absorption by hydrogen molecules in hydrogen atmosphere WDs cooler than about 4500~K causes infrared colors to become bluer with decreasing temperature, and begins to affect optical colors below about 4000~K, while helium atmosphere WDs of the same temperature have optical and infrared energy distributions similar to blackbodies.  Thus the addition of infrared data, or higher quality optical data, can help distinguish between atmosphere models for WDs beyond the turnover.  Figure~\ref{fig-gri} plots $g - r$ versus $r -i$ for stars with $\mbol > 16$ ($\teff \ga 3870$), with hydrogen and helium atmosphere evolutionary tracks overplotted.  Stars with preferred hydrogen and helium atmosphere models are plotted in cyan and magenta, respectively.  At $\teff \sim 3500$, the hydrogen and helium atmosphere evolutionary tracks separate by about 0.2 mag in $r - i$.  For these stars in our sample the SDSS photometric error is of order 0.1 mag, and thus distinguishing between atmosphere models is not possible for most of them.  Deeper photometry may help, though would require that other sources of error, such as in the extinction determination or atmosphere models themselves, be understood at the few percent level.  Much better leverage is obtained by adding infrared data.  \citet[][hereafter D16]{dame2016} obtained $J$ and $H$ photometry for 40 cool WDs selected from our survey, and fit hydrogen and helium atmosphere models to the combined infrared and SDSS photometry.  Figure~\ref{fig-rjh} plots their infrared colors for 11 stars with $\mbol > 15.5$ in our survey, again with hydrogen and helium atmosphere model cooling tracks overplotted.  Three of the stars were classified by D16 as pure hydrogen atmosphere WDs, indicated in cyan in Figure~\ref{fig-rjh}.  One was classified as having a mixed hydrogen and helium atmosphere (yellow in the figure).  The remaining seven were classified as having pure helium atmospheres (magenta in the figure), however D16 state that for those objects, the differences between the hydrogen and helium atmosphere model fits are small.  We thus consider those classified as helium atmosphere WDs to be better considered as not having a preferred atmosphere composition.  The overall impression of Figure~\ref{fig-rjh} is that it is consistent with our high adopted fraction of hydrogen atmosphere WDs past the turnover, and that clearly more accurate infrared photometry has the potential to allow unambiguous classification of atmospheric composition for stars cooler than $\teff \sim 4000~{\rm K}$.

\citet{kilic2010a} specifically addressed the problem of unknown atmospheric composition of cool WDs by obtaining follow-up $JHK$ photometry of most of the H06 WD sample with $\mbol > 14.6$.  They find 48\%, 35\%, and 17\% of their sample of 126 cool WDs have pure hydrogen, pure helium, and mixed hydrogen/helium atmospheres, respectively.  They found no pure helium atmosphere WDs cooler than 4500~K ($\mbol \sim 15.3$), and mostly mixed atmospheres cooler than 4000~K ($\mbol \sim 15.9$).  Their results thus support the low fraction of helium WDs we've adopted (see also \citealt{gianninas2015}).

\replaced{Perhaps a}{A} larger source of bias in interpreting the WDLF turnover is the unknown mass of most faint WDs.  Few intrinsically faint WDs have parallax measurements.  \citet{gianninas2015} obtained parallaxes for 54 cool WDs, and all six of their ultracool WDs ($\teff < 4000~{\rm K}$) have masses less than 0.4~$M_\sun$, versus the typical mass for hotter WDs of $\sim 0.6~M_\sun$ assumed in our analysis.  A 4000~K pure hydrogen atmosphere WD with a mass of 0.3~$M_\sun$ is $\sim 0.7$ mag brighter than a 0.6~$M_\sun$ WD of the same temperature and atmospheric composition.
Clearly \replaced{GAIA}{Gaia} will assist greatly in resolving this issue.

\subsection{Halo Results}

\begin{figure}
\plotone{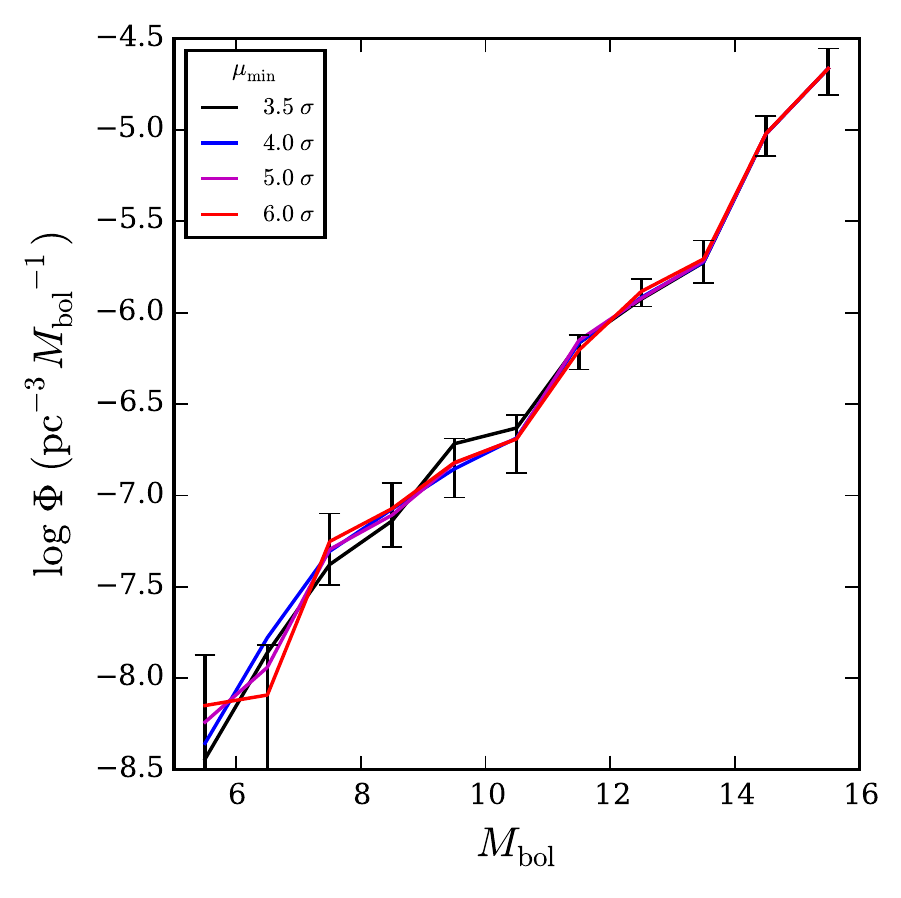}
\caption{LFs for our $200 < \vtan < 500\ \kmsec$ halo sample, but using different lower proper motion cuts (3.5, 4, 5, and 6~$\sigma$ for the black, blue, magenta, and red curves, respectively).  Error bars are for the $\mu_{\rm min} = 6.0~\sigma$ sample.}
\label{fig-halo-pm}
\end{figure}

\begin{figure}
\plotone{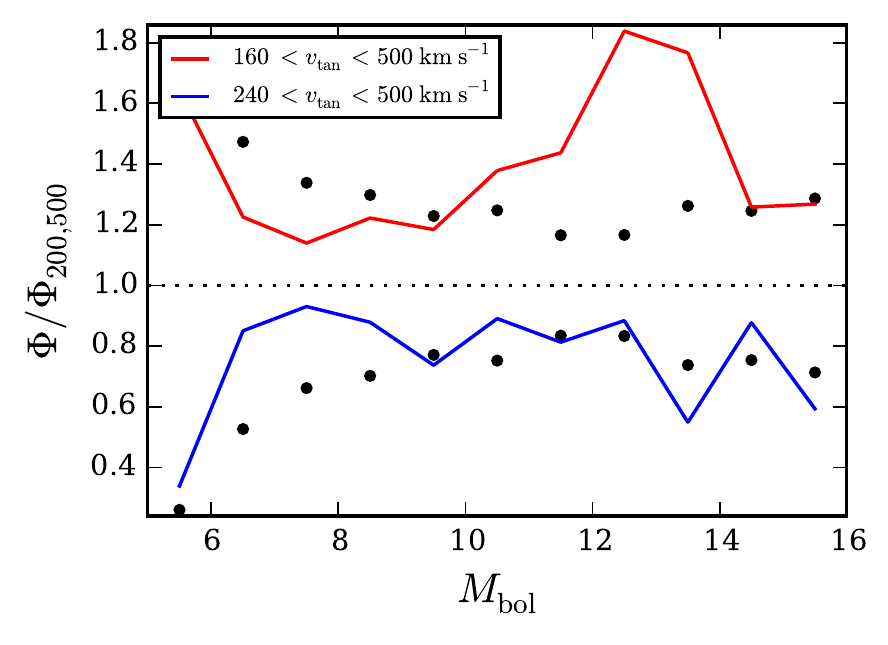}
\caption{Effect on the halo LF of different cuts in $\vtan$.  Colored lines indicate the ratio of the LF using alternate $\vtan$ cuts to our preferred LF ($200 < \vtan < 500\ \kmsec$).  The dots indicate the size of the Poisson errors in our preferred LF.}
\label{fig-halo-vtan}
\end{figure}

It is typical to isolate halo stars by limiting the sample of stars to those with $\vtan > 200\ \kmsec$.  This is consistent with the expected halo fraction versus $\vtan$ curve in Figure~\ref{fig-components}, and we adopt it for this work.  We also require $\vtan < 500\ \kmsec$ to remove unbound stars from the sample (e.g., \citealt{piffl2014} derive a local Galactic escape velocity of $533_{-41}^{+54}\ ~\kmsec$); this removes only 4.5 weighted stars.   Figure~\ref{fig-halo-pm} plots the halo LF for the sample of stars with $200 < \vtan < 500\ \kmsec$ for different lower proper motion limits.  Using 3.5, 4, 5, and 6~$\sigma$ lower proper motion cuts yields samples of 135, 124, 107, and 94 stars, respectively.  All four LFs agree within the errors.  Since there is no apparent significant contamination in the larger 3.5~$\sigma$ sample, we adopt it as our preferred halo sample.  Figure~\ref{fig-halo-vtan} displays the ratio of LFs using $\vtan$ cuts of $160 < \vtan < 500\ \kmsec$ and $240 < \vtan < 500\ \kmsec$ to our preferred LF with $200 < \vtan < 500\ \kmsec$.  Some contamination from the disk is apparent in the lower $\vtan > 160\ \kmsec$ cut.

\begin{figure}
\plotone{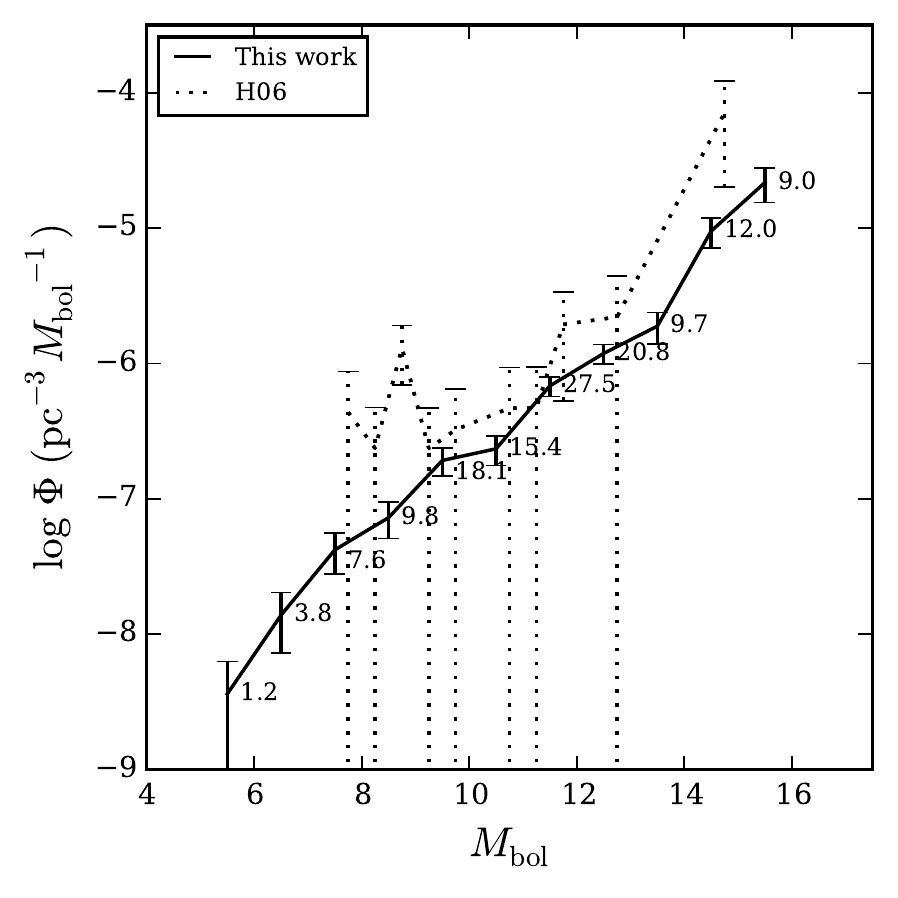}
\caption{{\it Solid line:} WDLF for our preferred halo sample ($200 < \vtan < 500\ \kmsec$, $\mu > 3.5~\sigma$).  Luminosity bins are 1 mag wide from $5.0 < \mbol < 16.0$.  Numbers next to each point are the number of weighted stars in that magnitude bin.  {\it Dashed line:} Halo WDLF from H06. (The LFs are available as the Data behind the Figure)}.
\label{fig-halo-lf-harris}
\end{figure}

\begin{figure}
\plotone{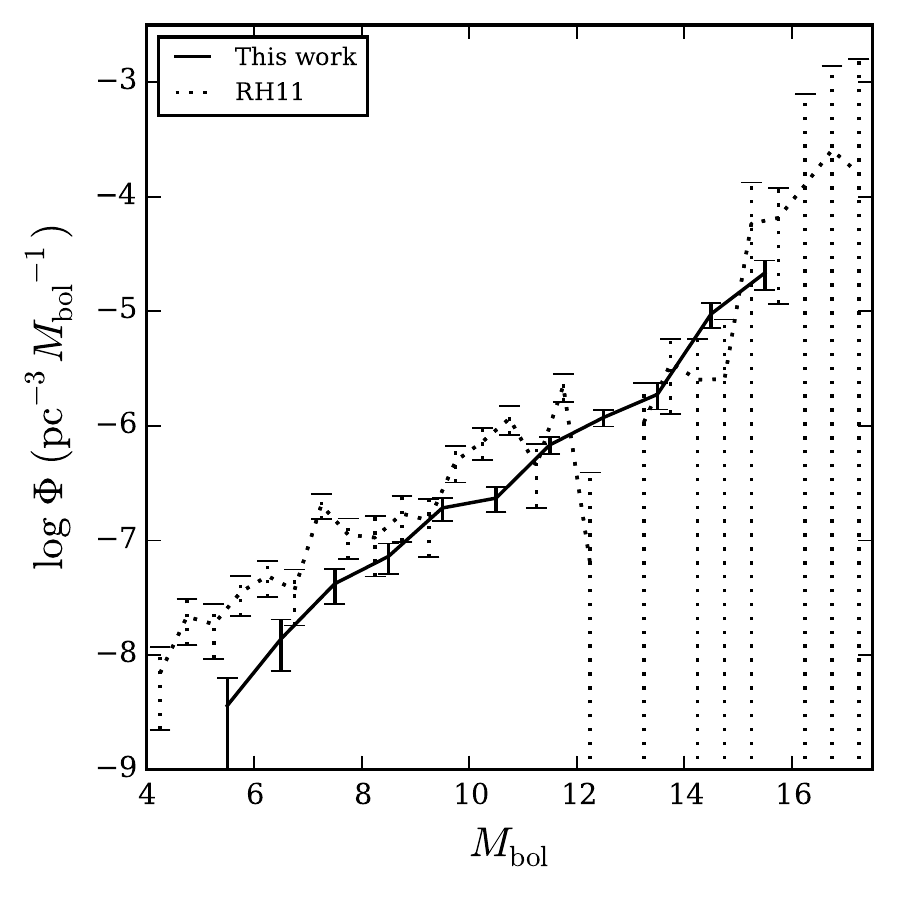}
\caption{{\it Solid line:} WDLF for our preferred halo sample ($200 < \vtan < 500\ \kmsec$, $\mu > 3.5~\sigma$, same as in Figure~\ref{fig-halo-lf-harris}).  {\it Dashed line:} Halo WDLF from RH11, scaled up by a factor of 2.00. (The LFs are available as the Data behind the Figure)}
\label{fig-halo-lf-rh}
\end{figure}

Our preferred halo sample ($\mu > 3.5~\sigma$, $200 < \vtan < 500\ \kmsec$) contains 135 stars, versus 18 and 93 with $\vtan > 200\ \kmsec$ for H06 and RH11, respectively.  Figures~\ref{fig-halo-lf-harris} and \ref{fig-halo-lf-rh} display our preferred halo LF, along with those of H06 ($\vtan > 200\ \kmsec$ LF from their Figure~10) and RH11 (their Figure~18), respectively.  The RH11 LF has been scaled up by the same factor of 2.00 we used to scale up their disk LF.    The turnover of the halo LF remains undetected, and will require deeper surveys to define.  The H06 LF agrees with ours within their error bars.  We find a total space density of halo WDs of $3.5 \pm 0.7 \times 10^{-5}$, consistent with H06's value of $4 \times 10^{-5}\ {\rm pc}^{-3}$.  RH11 find a larger overall density.  Within the luminosity range of their LF with the best statistics, $5 < \mbol < 12$, their WD density is on average 40\% larger than ours (after scaling to match the disk LFs).

\begin{figure}
\plotone{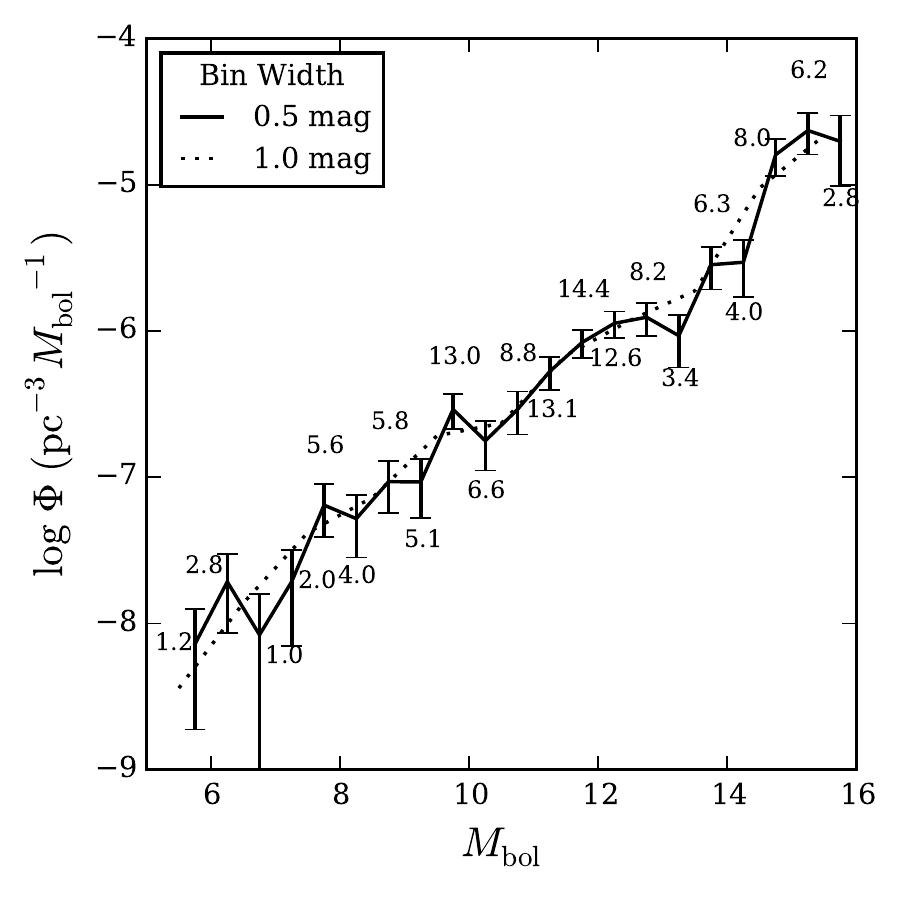}
\caption{Preferred halo LF with 0.5 mag wide bins (solid line, with numbers indicating the number of weighted stars in each bin), compared to LF with 1.0 mag wide bins (dotted line). (The LFs are available as the Data behind the Figure)}
\label{fig-halo-detail}
\end{figure}

With finer binning, we see the bump due to the onset of fully convective envelopes in the halo LF, just as we saw in the disk LF (Figure~\ref{fig-disk-detail}).  Figure~\ref{fig-halo-detail} displays the halo LF with 0.5 mag wide bins.  The rise occurs at the same luminosity as in the disk LF, though not as well defined due to the smaller number of stars.

\begin{figure}
\plotone{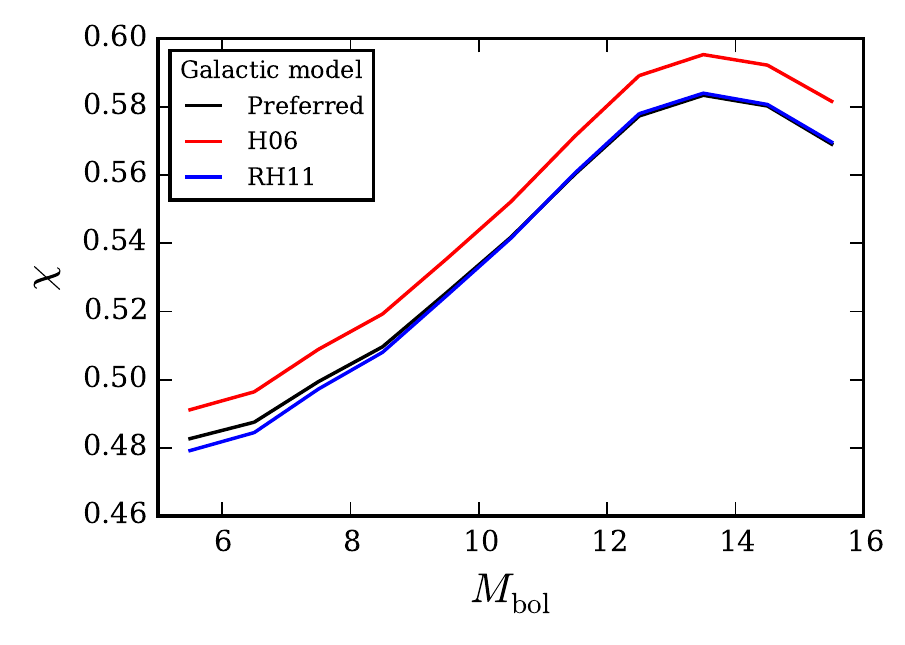}
\caption{Mean \deleted{volume-weighted }discovery fraction, $\chi$, versus $\mbol$ for our preferred halo WD sample, using different Galactic kinematic and density models.  Luminosity bins are the same as in Figure~\ref{fig-halo-lf-harris}.}
\label{fig-halo-df}
\end{figure}

Figure~\ref{fig-halo-df} displays the mean \deleted{volume-weighted }discovery fraction for our preferred halo WD sample, using different Galactic models.  The discovery fraction averaged over the entire sample for our preferred model is 0.55.  Similarly to the disk LF, the discovery fraction thus requires a large correction to the halo LF.  It is much less sensitive to the Galactic kinematic model than the disk discovery fraction is.  H06 used the halo velocity ellipsoid from \citet{morrison1990}, with dispersions ($\sigma_{v_R}$, $\sigma_{v_\phi}$, $\sigma_{v_z}$) of (133, 98, 94) $\kmsec$ and a rotation velocity relative to the Sun of $-206~\kmsec$.  RH11 used the halo velocity ellipsoid from \citet{chiba2000},  with dispersions ($\sigma_{v_R}$, $\sigma_{v_\phi}$, $\sigma_{v_z}$) of (141, 106, 94) $\kmsec$ and a rotation velocity relative to the sun of $-199~\kmsec$.  We use the inner halo velocity ellipsoid from \citet{carollo2010},  with dispersions ($\sigma_{v_R}$, $\sigma_{v_\phi}$, $\sigma_{v_z}$) of (150, 95, 85) $\kmsec$ and a rotation velocity relative to the sun of $-232~\kmsec$.  Differences between the models of order 10\% yield differences in the discovery fraction of only about 1\%, and with no significant trends with bolometric luminosity.

\begin{figure}
\plotone{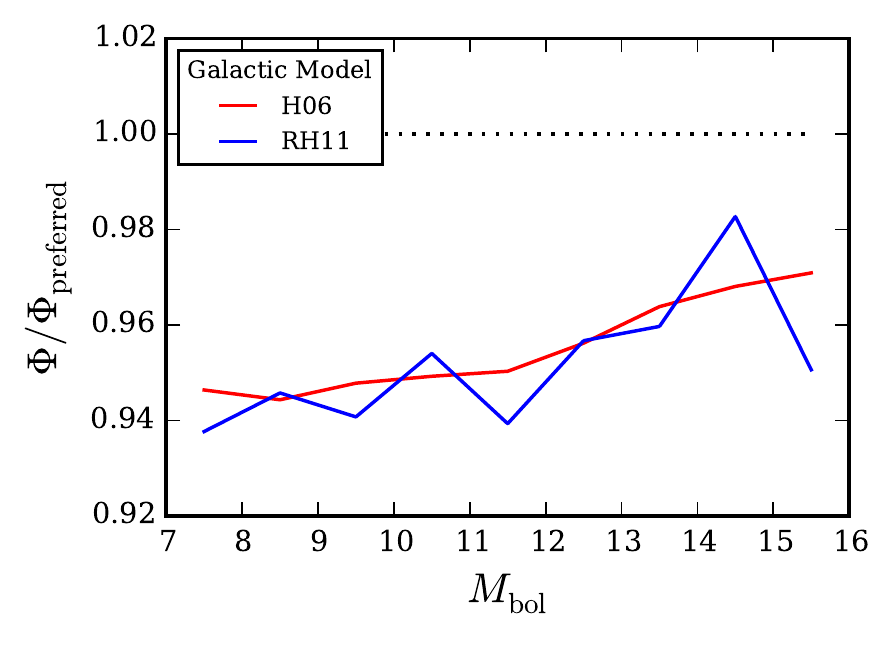}
\caption{Effect on the halo LF of different Galactic density and kinematic models.  Colored lines indicate the ratio of the LF using alternate Galactic models to our preferred LF.  The brightest two bins of the LF have been excluded due to small number statistics.}
\label{fig-halo-model}
\end{figure}

The halo LF is also far less sensitive to the choice of Galactic density model than the disk LF is.  This is primarily due to the fact that the halo density varies little over the local volume surveyed.  Figure~\ref{fig-halo-model} displays the ratio of our LFs using the H06 and RH11 models to the LF using our preferred model.  H06 and RH11 both assumed the halo density is constant over the survey volume.  This has the effect of lowering their halo densities by roughly 5\% compared to the \citet{juric2008} halo density model we adopted, smaller than the Poisson errors in any of our LFs.  There is a slight trend in the model differences with luminosity, of order 2\%.

\section{Catalog}
\label{section-catalog}

Table~\ref{table-stars} lists our catalog of 8472 WD candidates with $\mu > 3.5~\sigma$, $\vtan > 20\ \kmsec$, and an acceptable fit to either the pure hydrogen or pure helium atmosphere models.  The minimum cut in $\vtan$ is based on location within the RPM diagram (Figure~\ref{fig-rpm}), not the actual model fits.  Thus, there are some objects in the catalog for which the atmospheric model fits yield tangential velocities of less than $20~\kmsec$.  Positions and $ugriz$ magnitudes are listed from SDSS Data Release 7, and proper motions from M2014.  While $\chi^2$ values from both the hydrogen and helium atmosphere model fits are listed for all candidates, the actual fitted parameters are listed only for those fits deemed acceptable.

Tables~\ref{table-disk}, \ref{table-disk-big}, and \ref{table-halo} give the data necessary to construct LFs for three different samples: 1) our preferred disk sample, with $\mu > 5~\sigma$ and $40 < \vtan < 120~\kmsec$; 2) a disk sample with $\mu > 3.5~\sigma$ and $30 < \vtan < 120~\kmsec$, which yields a considerably larger sample size than our preferred disk sample but does suffer from contamination by subdwarfs at the faint end of the LF; and 3) our preferred halo sample, with $\mu > 3.5~\sigma$ and $200 < \vtan < 500~\kmsec$.  Only stars contained within each sample are listed in the respective tables.  The data given for each candidate WD includes the modified maximum survey volumes, discovery fractions, and probabilities that the star belongs to the targeted Galactic component, assuming both hydrogen and helium atmospheres, as well as the weights assigned to the hydrogen and helium atmosphere fits.  All values use our preferred Galactic model.  Our LFs thus may be reproduced, or modified with different binning, hydrogen/helium atmosphere model weights, and likelihoods of belonging to different Galactic populations.  Table~\ref{table-fields} lists each survey field, with the complete information required to calculate LFs using different $\vtan$ and proper motion cuts than were used in the paper, including the coordinates of each field center, the solid angle they cover on the sky, and the minimum proper motion cuts in each of their 0.1 mag wide $r$ bins.

\begin{splitdeluxetable*}{rrrrrrrrrrrrBrrrrrrrrrrrrrrrrBrrrrrrrrrrrrrrrr}
\tablecaption{WD Candidates\label{table-stars}}
\tablehead{
\multicolumn{12}{c}{} &
\multicolumn{10}{c}{} &
\multicolumn{5}{c}{Used in fits\tablenotemark{h}} &
\multicolumn{1}{c}{} &
\multicolumn{8}{c}{Hydrogen Atmosphere\tablenotemark{j}} &
\multicolumn{8}{c}{Helium Atmosphere\tablenotemark{j}} \\
\cmidrule(lr){23-27}
\cmidrule(lr){29-36}
\cmidrule(lr){37-44}
\colhead{ObjID\tablenotemark{a}} &
\colhead{Night\tablenotemark{b}} &
\colhead{ObsID\tablenotemark{c}} &
\colhead{CCD\tablenotemark{d}} &
\colhead{$\alpha$ (deg)} &
\colhead{$\delta$ (deg)} &
\multicolumn{2}{c}{$u$} &
\multicolumn{2}{c}{$g$} &
\multicolumn{2}{c}{$r$} &
\multicolumn{2}{c}{$i$} &
\multicolumn{2}{c}{$z$} &
\multicolumn{2}{c}{$\mu_\alpha\ ({\rm mas\ year^{-1}})$\tablenotemark{e}} &
\multicolumn{2}{c}{$\mu_\delta\ ({\rm mas\ year^{-1}})$} &
\colhead{$\mu_{\rm cut}$\tablenotemark{f}} &
\colhead{Comp\tablenotemark{g}} &
\colhead{u} &
\colhead{g} &
\colhead{r} &
\colhead{i} &
\colhead{z} &
\colhead{DOF\tablenotemark{i}} &
\colhead{$\chi^2$} &
\multicolumn{2}{c}{$T_{\rm eff}$ (K)} &
\multicolumn{2}{c}{d (pc)} &
\multicolumn{2}{c}{$M_{\rm bol}$} &
\colhead{$E(B\!\!-\!\!V)$} &
\colhead{$\chi^2$} &
\multicolumn{2}{c}{$T_{\rm eff}$ (K)} &
\multicolumn{2}{c}{d (pc)} &
\multicolumn{2}{c}{$M_{\rm bol}$} &
\colhead{$E(B\!\!-\!\!V)$}
}
\startdata
587722952767242621 & 54245 &  49 & 3 & 236.693017 &  -0.011476 & 24.958 &  0.797 & 22.821 &  0.141 & 21.143 &  0.049 & 20.520 &  0.047 & 20.003 &  0.115 &   -43.7 &  7.4 &   -60.5 &  7.6 &   6.74 & 0.923 & 0 & 1 & 1 & 1 & 1 & 2 &  25.92 & \nodata & \nodata & \nodata & \nodata & \nodata & \nodata & \nodata &   2.65 &   3500 &    53 &   66.0 &  14.3 & 16.440 &  0.471 & 0.010 \\
587722952768225477 & 55333 &  18 & 3 & 239.010919 &  -0.122741 & 19.522 &  0.029 & 19.441 &  0.016 & 19.728 &  0.020 & 19.906 &  0.033 & 20.135 &  0.135 &   -30.8 &  4.0 &     2.5 &  3.4 &   5.37 & 0.909 & 1 & 1 & 1 & 1 & 1 & 3 &   2.54 &  22470 &   738 &  534.0 & 120.5 &  8.254 &  0.539 & 0.131 &  12.04 & \nodata & \nodata & \nodata & \nodata & \nodata & \nodata & \nodata \\
587722952768618871 & 55333 &  18 & 4 & 239.853096 &  -0.156805 & 20.463 &  0.048 & 20.026 &  0.020 & 20.092 &  0.029 & 20.216 &  0.038 & 20.318 &  0.142 &    -1.5 &  3.7 &   -40.1 &  4.1 &   7.01 & 0.906 & 1 & 1 & 1 & 1 & 1 & 3 &   2.02 &  12692 &   725 &  414.2 &  85.6 & 10.774 &  0.544 & 0.126 &  20.86 & \nodata & \nodata & \nodata & \nodata & \nodata & \nodata & \nodata \\
587722952768684170 & 53890 &  12 & 1 & 240.033640 &  -0.144565 & 19.146 &  0.025 & 18.661 &  0.015 & 18.617 &  0.016 & 18.657 &  0.015 & 18.740 &  0.040 &   -33.9 &  3.6 &    28.1 &  3.5 &   8.36 & 0.932 & 1 & 1 & 1 & 1 & 1 & 3 &   2.08 &   8790 &   224 &  147.3 &  33.0 & 12.388 &  0.529 & 0.043 &  14.13 & \nodata & \nodata & \nodata & \nodata & \nodata & \nodata & \nodata \\
587722952769667554 & 53890 &  14 & 1 & 242.285879 &  -0.075495 & 22.284 &  0.195 & 21.205 &  0.035 & 20.823 &  0.037 & 20.688 &  0.049 & 21.117 &  0.268 &    26.2 &  7.2 &   -28.6 &  7.2 &   4.04 & 0.925 & 1 & 1 & 1 & 1 & 0 & 2 &   2.15 &   6197 &   162 &  227.4 &  44.9 & 13.921 &  0.469 & 0.063 &   1.95 &   6232 &   182 &  229.0 &  47.9 & 13.924 &  0.482 & 0.063 \\
\enddata
\tablecomments{Table~\ref{table-stars} is published in its entirety in machine-readable format.  A portion is shown here for guidance regarding its form and content.}\tablecomments{Positions and $ugriz$ photometry are from the SDSS Data Release 7.  Proper motions are from M2014.}
\tablenotetext{a}{Unique identifier in SDSS Data Release 7.  Corresponds to the {\it objID} column of Table~4 of M2014.}
\tablenotetext{b}{MJD number of the night the observation was obtained in M2014.  Corresponds to the {\it Night} column in both Table~\ref{table-fields} in this paper as well as Table~2 of M2014.}
\tablenotetext{c}{Observation number in M2014, unique within a given night.  Corresponds to the {\it ObsID} column in both Table~\ref{table-fields} in this paper as well as Table~2 of M2014.}
\tablenotetext{d}{CCD on which the object was detected in M2014.  Corresponds to the {\it CCD} column of Table~\ref{table-fields}.}
\tablenotetext{e}{$\dot{\alpha}\cos(\delta)$}
\tablenotetext{f}{Total proper motion expressed as a multiple of the estimated proper motion error in its subsample.}
\tablenotetext{g}{Correction factor for survey completeness.}
\tablenotetext{h}{1 if magnitude was used in model fits, 0 if not used.}
\tablenotetext{i}{Degrees of freedom in model fits.}
\tablenotetext{j}{$\chi^2$ values are listed for all model atmosphere fits.  The fit parameters are listed only if the fit is considered acceptable.}
\end{splitdeluxetable*}

\begin{deluxetable*}{rrrrrrrrr}
\tablecaption{WD Candidates in Preferred Disk Sample ($\mu > 5.0, 40 < \vtan < 120\ {\rm km~s^{-1}})$\label{table-disk}}
\tablehead{
\multicolumn{1}{c}{} &
\multicolumn{4}{c}{Hydrogen Atmosphere\tablenotemark{b}} &
\multicolumn{4}{c}{Helium Atmosphere\tablenotemark{b}} \\
\cmidrule(lr){2-5}
\cmidrule(lr){6-9}
\colhead{ObjID\tablenotemark{a}} &
\colhead{Weight\tablenotemark{c}} &
\colhead{Prob\tablenotemark{d}} &
\colhead{$V_{\rm mod}\ ({\rm pc^3})$} &
\colhead{$\chi$\tablenotemark{e}} &
\colhead{Weight\tablenotemark{c}} &
\colhead{Prob\tablenotemark{d}} &
\colhead{$V_{\rm mod}\ ({\rm pc^3})$} &
\colhead{$\chi$\tablenotemark{e}}
}
\startdata
587722952768225477 & 1.000 & 0.997 &   2330815.4 & 0.327 & 0.000 & \nodata & \nodata & \nodata \\
587722952768618871 & 1.000 & 0.998 &   1834089.7 & 0.343 & 0.000 & \nodata & \nodata & \nodata \\
587722952772092251 & 0.685 & 1.000 &    763030.9 & 0.364 & 0.315 & 1.000 &    780805.3 & 0.363 \\
587722953304309938 & 0.000 & \nodata & \nodata & \nodata & 1.000 & 1.000 &   1334854.9 & 0.354 \\
587722953305882673 & 1.000 & 0.988 &   1949106.5 & 0.340 & 0.000 & \nodata & \nodata & \nodata \\
\enddata
\tablecomments{Table~\ref{table-disk} is published in its entirety in machine-readable format.  A portion is shown here for guidance regarding its form and content.}\tablenotetext{a}{Unique identifier in SDSS Data Release 7.}
\tablenotetext{b}{Values are listed only if atmospheric model has an acceptable fit.}
\tablenotetext{c}{Weight assigned to this atmospheric model.}
\tablenotetext{d}{Probability star belongs to the targeted Galactic component.}
\tablenotetext{e}{\replaced{Volume-weighted discovery}{Discovery} fraction.}
\end{deluxetable*}

\begin{deluxetable*}{rrrrrrrrr}
\tablecaption{WD Candidates in Contaminated Disk Sample ($\mu > 3.5, 30 < \vtan < 120\ {\rm km~s^{-1}})$\label{table-disk-big}}
\tablehead{
\multicolumn{1}{c}{} &
\multicolumn{4}{c}{Hydrogen Atmosphere} &
\multicolumn{4}{c}{Helium Atmosphere} \\
\cmidrule(lr){2-5}
\cmidrule(lr){6-9}
\colhead{ObjID} &
\colhead{Weight} &
\colhead{Prob} &
\colhead{$V_{\rm mod}\ ({\rm pc^3})$} &
\colhead{$\chi$} &
\colhead{Weight} &
\colhead{Prob} &
\colhead{$V_{\rm mod}\ ({\rm pc^3})$} &
\colhead{$\chi$}
}
\startdata
587722952768225477 & 1.000 & 0.997 &   5557623.6 & 0.436 & 0.000 & \nodata & \nodata & \nodata \\
587722952768618871 & 1.000 & 0.998 &   4048229.7 & 0.461 & 0.000 & \nodata & \nodata & \nodata \\
587722952768684170 & 1.000 & 1.000 &   2551750.7 & 0.483 & 0.000 & \nodata & \nodata & \nodata \\
587722952769667554 & 0.686 & 1.000 &   1179623.6 & 0.507 & 0.314 & 1.000 &   1201655.7 & 0.507 \\
587722952771568497 & 1.000 & 1.000 &   3030914.2 & 0.476 & 0.000 & \nodata & \nodata & \nodata \\
\enddata
\tablecomments{Table~\ref{table-disk-big} is published in its entirety in machine-readable format.  A portion is shown here for guidance regarding its form and content.}\end{deluxetable*}

\begin{deluxetable*}{rrrrrrrrr}
\tablecaption{WD Candidates in Preferred Halo Sample ($\mu > 3.5, 200 < \vtan < 500\ {\rm km~s^{-1}})$\label{table-halo}}
\tablehead{
\multicolumn{1}{c}{} &
\multicolumn{4}{c}{Hydrogen Atmosphere} &
\multicolumn{4}{c}{Helium Atmosphere} \\
\cmidrule(lr){2-5}
\cmidrule(lr){6-9}
\colhead{ObjID} &
\colhead{Weight} &
\colhead{Prob} &
\colhead{$V_{\rm mod}\ ({\rm pc^3})$} &
\colhead{$\chi$} &
\colhead{Weight} &
\colhead{Prob} &
\colhead{$V_{\rm mod}\ ({\rm pc^3})$} &
\colhead{$\chi$}
}
\startdata
587722982832734535 & 0.907 & 1.000 & 184188380.0 & 0.502 & 0.093 & 1.000 & 173244465.4 & 0.504 \\
587722982836928948 & 1.000 & 1.000 & 130084347.1 & 0.515 & 0.000 & \nodata & \nodata & \nodata \\
587722983357087771 & 1.000 & 1.000 &  89714036.8 & 0.532 & 0.000 & \nodata & \nodata & \nodata \\
587722983890354441 & 0.000 & 1.000 & 196251208.9 & 0.500 & 0.084 & 1.000 & 164044409.2 & 0.506 \\
587722983900512884 & 0.773 & 1.000 &    376128.8 & 0.566 & 0.227 & 1.000 &    348782.2 & 0.564 \\
\enddata
\tablecomments{Table~\ref{table-halo} is published in its entirety in machine-readable format.  A portion is shown here for guidance regarding its form and content.}\end{deluxetable*}

\begin{deluxetable*}{rrrrrrrrrrrrr}
\tablecaption{Survey Fields\label{table-fields}}
\tablehead{
\multicolumn{6}{c}{} &
\multicolumn{7}{c}{$\sigma_\mu\ ({\rm mas~year^{-1}})$\tablenotemark{c}} \\
\cline{7-13}
\colhead{Night\tablenotemark{a}} &
\colhead{ObsID\tablenotemark{b}} &
\colhead{CCD} &
\colhead{$\alpha$ (deg)} &
\colhead{$\delta$ (deg)} &
\colhead{$\Omega\ ({\rm deg^2})$} &
\colhead{16.0} &
\colhead{16.1} &
\colhead{16.2} &
\colhead{16.3} &
\colhead{16.4} &
\colhead{16.5} &
\colhead{16.6}
}
\startdata
53737 &  27 & 1 & 118.203423 &  28.214987 & 0.170 & 27.35 & 27.35 & 27.35 & 27.36 & 27.36 & 27.36 & 27.36 \\
53737 &  27 & 2 & 118.203423 &  28.214987 & 0.159 & 23.65 & 23.65 & 23.65 & 23.65 & 23.66 & 23.66 & 23.66 \\
53737 &  27 & 3 & 118.203423 &  28.214987 & 0.168 & 11.61 & 11.61 & 11.61 & 11.61 & 11.61 & 11.62 & 11.62 \\
53737 &  27 & 4 & 118.203423 &  28.214987 & 0.182 & 18.08 & 18.08 & 18.08 & 18.08 & 18.08 & 18.08 & 18.08 \\
53740 & 151 & 1 &  29.481613 &  14.290259 & 0.168 &  7.27 &  7.27 &  7.28 &  7.28 &  7.28 &  7.28 &  7.28 \\
\enddata
\tablecomments{Table~\ref{table-fields} is published in its entirety in machine-readable format.  A portion is shown here for guidance regarding its form and content.  There are an additional 48 columns specifying the proper motion errors in fainter bins.}\tablenotetext{a}{MJD number of the night the observation was obtained in M2014.  Corresponds to the {\it night} column in Table~2 of M2014.}
\tablenotetext{b}{Observation number in M2014, unique within a given night.  Corresponds to the {\it obsID} column in Table~2 of M2014.}
\tablenotetext{c}{Error in proper motion for each 0.1 mag wide $r$ bin (i.e., subsample), used to set the minimum proper motion in that bin.  Each column is marked with the brighter $r$ limit for that bin.  The last two bins for 1.3m fields will not contain data, as they are beyond the survey limits for the 1.3m.}
\end{deluxetable*}


\section{Summary}
\label{section-summary}

We have presented a reduced-proper-motion selected sample of 8472 WD candidates from a deep proper motion catalog, covering 2256 square degrees of sky to faint $r$ limits of 21.3 -- 21.5.  SDSS $ugriz$ photometry has been fit to pure hydrogen and helium atmosphere model WDs to derive distances, bolometric luminosities, effective temperatures, and atmospheric compositions.  The disk WDLF has been presented, with statistically significant samples of stars cooler than the LF turnover, a region of the LF particularly important in applying the WDLF to the determination of the age of the disk.  That determination remains hampered by the unknown atmospheric composition and masses of most stars fainter than the turnover, and likely won't be resolved until \replaced{GAIA}{Gaia} parallaxes for significant samples of faint WDs are available.  The halo WDLF has also been presented, based on a sample of 135 stars.  Both the disk and halo WDLFs have been compared to those of H06 and RH11, which are similar in technique, the portion of the LF studied, and sample size.  The shape of the disk LF is in broad agreement with both H06 and RH11 brighter than the turnover.  We find a higher density of WDs fainter than the turnover, though only at the 2 -- 3 $\sigma$ level.   Our halo WDLF agrees with the H06 LF within their errors, but the RH11 LF gives densities about 40\% larger (after scaling to match our disk LFs).  The turnover in the halo WDLF remains undetected.  We detect with high statistical significance the bump in the disk LF due to the onset of fully convective envelopes in WDs, a feature first seen by H06, and see indications of it in the halo LF as well.

While these are the largest samples to date of disk WDs fainter than the turnover, as well as of halo WDs, the imminent release of the the Panoramic Survey Telescope and Rapid Response System 3pi survey data will allow for both deeper and much larger samples, with improved photometric accuracy that should help in distinguishing between hydrogen and helium atmosphere WDs for those WDs cooler than the turnover in the disk LF.  In about a year we anticipate the revolution in WD research that Data Release 2 of \replaced{GAIA}{Gaia} will offer, and in just a few years the second revolution that the Large Synoptic Survey Telescope will deliver.  In a field that has long been challenged by insufficiently small samples, the next few years are going to be rewarding.

\acknowledgements
We thank N. Rowell and N. C. Hambly for  providing us with the data for their luminosity functions.  \added{We also thank the referee, E. Garc{\'{i}}a-Berro, for helpful suggestions. }This material is based on work supported by the National Science Foundation under grant AST 06-07480.  MK, KD, and AG gratefully acknowledge the support of the NSF and NASA
under grants AST-1312678 and NNX14AF65G.  KW gratefully acknowledges the support of the NFS under grant AST-0602288.

Funding for the SDSS and SDSS-II has been provided by the Alfred
P. Sloan Foundation, the Participating Institutions, the National
Science Foundation, the U.S. Department of Energy, the National
Aeronautics and Space Administration, the Japanese Monbukagakusho, the
Max Planck Society, and the Higher Education Funding Council for
England. The SDSS Web Site is http://www.sdss.org/.
The SDSS is managed by the Astrophysical Research Consortium for the
Participating Institutions. The Participating Institutions are the
American Museum of Natural History, Astrophysical Institute Potsdam,
University of Basel, University of Cambridge, Case Western Reserve
University, University of Chicago, Drexel University, Fermilab, the
Institute for Advanced Study, the Japan Participation Group, Johns
Hopkins University, the Joint Institute for Nuclear Astrophysics, the
Kavli Institute for Particle Astrophysics and Cosmology, the Korean
Scientist Group, the Chinese Academy of Sciences (LAMOST), Los Alamos
National Laboratory, the Max-Planck-Institute for Astronomy (MPIA),
the Max-Planck-Institute for Astrophysics (MPA), New Mexico State
University, Ohio State University, University of Pittsburgh,
University of Portsmouth, Princeton University, the United States
Naval Observatory, and the University of Washington.

The Digitized Sky Surveys were produced at the Space Telescope Science Institute under U.S. Government grant NAG W-2166. The images of these surveys are based on photographic data obtained using the Oschin Schmidt Telescope on Palomar Mountain and the UK Schmidt Telescope. The plates were processed into the present compressed digital form with the permission of these institutions.

\facility{Bok(90prime), USNO:1.3m(Array Camera)}
\software{Astropy \citep{astropy}, Image
Reduction and Analysis Facility \citep[IRAF;][]{tody1986,
  tody1993}\footnote{IRAF is distributed by the National Optical
  Astronomy Observatories, which are operated by the Association of
  Universities for Research in Astronomy, Inc., under cooperative
  agreement with the National Science Foundation.}, DS9, SExtractor \citep{bertin1996},
  DAOPHOT II \citep{stetson1987}}

\end{document}